\documentclass[12pt]{article}
\usepackage{eurosym}
\usepackage{graphicx}
\usepackage{amsthm}
\usepackage{amsmath}
\usepackage{amssymb}
\usepackage{appendix}
\usepackage{color}
\usepackage{sgame}
\usepackage{verbatim}
\usepackage{caption}
\usepackage{subcaption}
\usepackage{mathtools}
\usepackage{dsfont}
\usepackage{float}
\usepackage[margin=1in]{geometry}
\usepackage{natbib}
\usepackage[utf8]{inputenc}
\usepackage[english]{babel}
\usepackage{lscape}
\usepackage[colorlinks=true,allcolors=blue]{hyperref}
\usepackage{booktabs}
\usepackage{longtable}
\usepackage{threeparttable}
\usepackage{dcolumn}
\usepackage{tikz}
\usepackage{xcolor}
\usepackage{istgame}
\usepackage{multirow}
\usepackage{amsfonts}
\usepackage{adjustbox}
\usepackage{istgame}
\usepackage{bbm}
\usepackage{soul}
\usepackage{dsfont}

\usetikzlibrary{calc}
\usetikzlibrary{calc}
\begin{document}

\title{Revisiting the Intra-Team Communication Method to Elicit Level-$k$ Reasoning in Beauty Contests and 11--20 Games\thanks{Financial support from the Qualitative Collaboration at the University of Virginia is gratefully acknowledged. We thank Charles A. Holt and Joseph Tao-yi Wang for their valuable comments and support. We are also grateful to Pin-Hsun Chen, Yan Ting Chen, Moritz Lakenbrink, and Linda Plehn for their excellent research assistance. Finally, we thank Konrad Burchardi and Stefan Penczynski for generously sharing their experimental data with us. The experiment is approved by the University of Virginia IRB \#7979.}}
\author{Zitian Wang\thanks{%
University of Virginia, Charlottesville, VA 22904. Email: 
zw3qt@virginia.edu} \and
Istiak Ahmed\thanks{%
University of Virginia, Charlottesville, VA 22904. Email: 
mqm3sk@virginia.edu} \and
Patarasate Unjitwattana\thanks{%
University of Virginia, Charlottesville, VA 22904. Email: 
amu8kw@virginia.edu} \and
Emily Yunxi Xie\thanks{%
University of Virginia, Charlottesville, VA 22904. Email: 
hky2sj@virginia.edu} \and
Meng-Jhang Fong\thanks{%
(Corresponding Author) Max Planck Institute for Behavioral Economics, Bonn, Germany 53113. Email: 
fong@econ.mpg.de} \and 
Po-Hsuan Lin\thanks{%
(Corresponding Author) University of Virginia, Charlottesville, VA 22904. Email: 
plin@virginia.edu} \and
}

\date{\today}

\maketitle

\begin{abstract}
How level-0 players behave and how they are perceived by higher-level players are central questions in the literature on level-$k$ models of boundedly rational strategic reasoning. To study these twin questions, we apply the \emph{intra-team communication method} developed by \cite{burchardi2014out} to identify level-0 actions and beliefs in the canonical beauty contest game and in a variant of the 11--20 game of \cite{goeree2018noisy}, in which behavior appears inconsistent with the standard level-$k$ model. In the beauty contest game, we replicate \cite{burchardi2014out}'s finding that elicited level-0 beliefs align with observed level-0 actions. In the variant of the 11--20 game, however, elicited level-0 beliefs and observed level-0 actions diverge, and both depart from the standard level-0 assumption, suggesting a complementary explanation for the behavioral pattern documented by \cite{goeree2018noisy}.

\end{abstract} 

\renewcommand{\baselinestretch}{1.2}

\thispagestyle{empty}

\medskip

{\small
\noindent JEL Classification Number: C81, C91, D83

\noindent Keywords: Cognitive Hierarchies, Communication, Replication, Experimental Methodology
}

\newpage\setcounter{page}{1}\renewcommand{\baselinestretch}{1.2}

\section{Introduction}\label{section:intro}

In behavioral game theory, the ``Level-$k$'' model \citep{nagel1995unraveling, stahl1995players} has become a workhorse solution concept for organizing widely observed non-equilibrium behavior in strategic environments.\footnote{Level-$k$ hierarchical reasoning has been observed in a wide variety of games, including, but not limited to, matrix games (e.g., \citealp{stahl1994experimental, stahl1995players, costa2001cognition, crawford2007fatal}), investment games (e.g., \citealp{rapoport2000mixed}), market-entry games (e.g., \citealp{camerer2004cognitive}), guessing games (e.g., \citealp{costa2006cognition}), sender–receiver games (e.g., \citealp{cai2006overcommunication, wang2010pinocchio, fong2023extreme}), auctions (e.g., \citealp{crawford2007level}), ring games (e.g., \citealp{kneeland2015identifying, chen2025measuring}), centipede games (e.g., \citealp{garcia2020non, lin2024cognitive, hu2026strategy}), and dirty-faces games (e.g., \citealp{lin2022cognitivemultii}).} Rather than assuming mutually consistent beliefs, the level-$k$ model posits a hierarchy of heterogeneous sophistication: level-0 players are non-strategic, while each higher-level player believes others are one level below them and best responds accordingly. As a result, the model's predictions hinge on how level-0 players' behavior is specified. However, both their actual behavior and strategic players' beliefs about them remain important open questions, motivating approaches that elicit reasoning beyond observed choices.

To this end, \citeauthor{burchardi2014out} (\citeyear{burchardi2014out}, henceforth BP) develop the \emph{intra-team communication method}, in which subjects are randomly and anonymously matched into two-person teams that split earnings equally. 
Each subject first submits a suggested decision along with a one-way, free-form message intended to ``convince'' their teammate to adopt that decision. After the messages are exchanged, both team members submit a final decision, one of which is implemented by the computer as the team's action with equal probability. Because the suggested decision and message are composed before a subject sees their teammate's input, they capture the subject's own reasoning rather than a reaction to their teammate's reasoning. Moreover, since the message is the only channel through which a subject can influence the team's action, subjects have an incentive to make their reasoning explicit.
This method represents a key innovation in the experimental study of strategic behavior: it provides one of the few approaches for directly eliciting subjects' underlying reasoning and has been widely applied across a range of domains.\footnote{The intra-team communication method has been applied in a variety of contexts, including beauty contest games \citep{burchardi2014out}, hide-and-seek games \citep{penczynski2016strategic}, persuasion \citep{penczynski2016persuasion}, social learning \citep{penczynski2017nature}, coordination games \citep{van2020coordination}, contests \citep{bruner2022strategic}, and the Colonel Blotto game \citep{arad2024does, arad2024multi}.}

BP first applied this method to the beauty contest game, a cornerstone of the level-$k$ literature, allowing them to estimate level-0 actions and beliefs empirically rather than impose ad hoc assumptions.\footnote{In a standard $p$-beauty contest game, each participant selects an integer between 0 and 100, and the participant whose choice is closest to $p$ times the average of all choices wins a prize. In BP, $p=2/3$.} Overall, BP identify about one-fifth of subjects as non-strategic players using their message data. Their actions are not uniformly distributed, while higher-level players' beliefs about non-strategic behavior are, on average, broadly consistent with the observed behavior. Despite the method's subsequent influence, the original beauty contest experiment itself has never been replicated, leaving open whether its findings represent a durable empirical regularity or merely reflect the behavior of a single subject pool at a particular time and place. In this paper, we provide the first replication of BP's beauty contest experiment, following their protocol exactly with a new subject pool more than a decade after the original study.

In the beauty contest game, we replicate BP's key findings: the distribution of elicited level-0 beliefs\footnote{Throughout, ``level-0 belief'' denotes a higher-level player's belief about level-0 players' actions, not a belief held by level-0 players themselves.} in our experiment is statistically indistinguishable from theirs, even though our subjects' suggested and final choices lie systematically below those in BP. As in BP, these elicited level-0 beliefs also align with the observed choices of subjects classified as level-0, replicating the belief–action consistency documented in their original study. Together, these results demonstrate that the intra-team communication method reliably recovers the reasoning underlying observed choices, motivating us to leverage this method to unpack the 11--20 puzzle documented by \citeauthor{goeree2018noisy} (\citeyear{goeree2018noisy}, henceforth GLZ).

The original 11--20 game, proposed by \cite{arad201211}, asks two players to simultaneously choose an integer between 11 and 20, with each player receiving an amount equal to their chosen number plus a bonus of 20 if that number is exactly one less than their opponent's. Since more money is better, it is natural to expect a non-strategic level-0 player to choose the highest amount, 20. Given this level-0 specification, the standard level-$k$ model predicts that each higher-level player undercuts the level below by one as a best response, so that levels 1 through 3 choose 19, 18, and 17, respectively.

\begin{figure}[htbp!]
    \centering
    \includegraphics[width=0.8\linewidth]{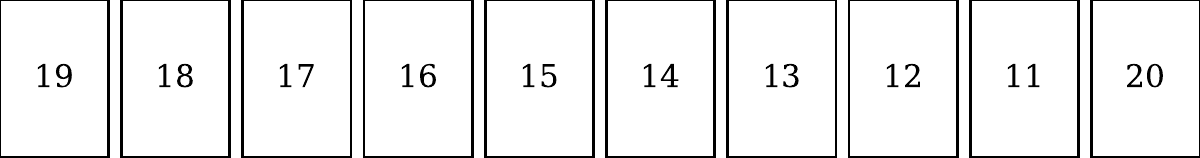}
    \caption{The Extreme version of the 11--20 game of \cite{goeree2018noisy}}
    \label{fig:intro_extreme_1120}
\end{figure}

GLZ test this logic by placing the ten integers into ten boxes arranged on a line, always keeping 20 in the rightmost box, and awarding the bonus to the player who chooses the box immediately to the left of their opponent's box, thereby preserving the undercutting logic under any reordering of the integers. In their \emph{Extreme} version (Figure~\ref{fig:intro_extreme_1120}), the integers 11 through 19 are arranged in \textit{descending} order from left to right. 
Because the chain of best responses operates over box positions rather than numerical labels, the standard level-$k$ model predicts that each level chooses the same box position in both the original and Extreme versions of the 11--20 game. 
Consequently, 19, the level-1 choice in the original game, becomes the level-9 choice in the Extreme version.

When playing the Extreme version, however, subjects concentrate their choices on the two end boxes, 19 and 20,\footnote{See Figure \ref{fig:main_1120_figure} for the distribution of choices.} with the prevalence of 19 implying an implausible jump to level~9, far above the reasoning levels typically observed in previous level-$k$ studies.
GLZ attribute this anomaly to common knowledge of noise, formalized by the noisy introspection model of \cite{goeree2004model}, in which subjects not only respond noisily to their own beliefs but also anticipate noisy responses by others at every level of reasoning.

In this paper, we provide the first replication of the Extreme 11--20 game and confirm GLZ's aggregate finding that choices concentrate on the two end boxes. The reasoning-process data elicited using the intra-team communication method let us examine level-0 beliefs and actions directly, mirroring our analysis of the beauty contest game. We find that both the elicited beliefs about level-0 play and the observed choices of subjects classified as level-0 differ significantly from the degenerate distribution at 20, the standard level-0 specification. Moreover, unlike in the beauty contest game, where level-0 beliefs and actions are consistent, here they diverge instead, providing the level-$k$ literature with an unusual case in which higher-level players systematically misperceive level-0 behavior---a conceptual replication of BP's belief-action comparison, but with the opposite result.
Since the level-$k$ predictions for the Extreme 11--20 game are more sensitive to the level-0 specification than those for the standard game, the non-standard level-0 actions and beliefs we identify offer a complementary explanation for the concentration on 19 and 20: whereas GLZ's noisy-introspection account focuses on noisy responses by higher-level players, ours highlights departures from standard assumptions about level-0 behavior and beliefs.

\section{Experimental Procedures and Message Classification}

\subsection{Experimental Design and Implementation}

We conduct a within-subject laboratory experiment in which each subject plays six one-shot games, organized into two blocks of three: three $p$-beauty contest games, with $p$ taking the values 2/3 (as in BP), 1/3, and 1/2, in that fixed order, and three 11--20 games corresponding to GLZ's \emph{Extreme}, \emph{Moderate}, and \emph{Basic} versions, also in that fixed order.\footnote{In the \emph{Basic} version, the ten integers appear in increasing order (11--20), matching the original game of \cite{arad201211}. The \emph{Moderate} version places 19 in the middle of the sequence (14, 13, 12, 11, 19, 18, 17, 16, 15, 20), whereas the \emph{Extreme} version reverses the order entirely (Figure \ref{fig:intro_extreme_1120}). In all three versions, 20 remains in the rightmost box.} The order of games within each block is identical for all subjects, whereas the order of the two blocks is counterbalanced across sessions to mitigate potential order effects. Although subjects receive no feedback across games, repeated play of similar games could still generate learning effects. We therefore focus our analysis on the first game of each block---the 2/3-beauty contest and the Extreme 11--20 game---to avoid this potential spillover. We relegate the analysis of the remaining four games to \ref{appendix:tables_figures}.

In each game, subjects are randomly matched into a new two-person team and follow BP's intra-team communication protocol (as described in Section \ref{section:intro}). The team's action, randomly selected from the two members' final decisions, determines their payoff. See \ref{appendix:instructions} for the experimental instructions and \ref{appendix:screenshots} for screenshots of the experimental interface.

We conducted six sessions with 12 subjects each ($N = 72$) at the Veconlab of the University of Virginia,\footnote{Specifically, subjects in three sessions played the three beauty contest games first, whereas subjects in the other three sessions played the three 11--20 games first.} recruiting participants from a subject pool open to all students. Following BP, each session began with two unpaid practice rounds in which teams guessed the year of a historical event, familiarizing subjects with the message-exchange procedure without inducing strategic reasoning. The instructions for each block were distributed only after the preceding block had concluded, and the experimental interface is programmed in oTree \citep{chen2016otree}. Sessions lasted approximately 45 minutes. Subjects were paid based on their earnings across all six games. 
Specifically, the winning team in each beauty contest game earned 100 points, while in each 11--20 game, a team received the number of points indicated by the selected box plus a 20-point bonus for successful undercutting.
At a conversion rate of 4 points to \$1, average earnings were \$24.1 (including a \$10 show-up fee), ranging from \$15.1 to \$46.5.

\subsection{Message Classification}
\label{subsec:message_classification}

We classify messages following BP's two-stage procedure. Four research assistants (RAs) familiar with the level-$k$ model are assigned in pairs to the two classes of games, one pair to the beauty contest messages and the other to the 11--20 messages. Each pair independently classifies the 216 messages (72 subjects $\times$ 3 games) for its assigned class of games.

In Stage 1, RAs see only the messages and record, for each subject, (i) a lower bound on steps of reasoning, defined as the minimum level-$k$ type consistent with the message, and (ii) a level-0 belief, defined as the non-derived belief about non-strategic play from which reasoning begins.\footnote{In the beauty contest games, the level-0 belief is recorded as a single number between 0 and 100, with stated ranges recorded as their midpoint, since only the mean of others' choices matters for best responses. In the 11--20 games, by contrast, best responses depend on the opponent team's full choice distribution. Accordingly, the level-0 belief is recorded as the specific integer or integers between 11 and 20 named in the message. When multiple integers are specified, we assume a uniform distribution over that support.} Withholding suggested choices at this stage prevents them from biasing the lower-bound classification. In Stage 2, after Stage 1 is complete, RAs additionally observe the suggested choices and record (i) an upper bound, defined as the maximum level-$k$ type consistent with both the message and the suggested choice, and (ii) two game-specific dummies capturing rationales that fall outside the level-$k$ framework but may coexist with it---equilibrium identification and dominance reasoning in the beauty contest games, following BP, and payoff security and undercut aversion in the 11--20 games.\footnote{Equilibrium identification records whether the subject identifies 0 as the equilibrium outcome, while dominance reasoning records the use of iterated deletion of dominated strategies. Payoff security records whether the subject considers securing a safe payoff when making a choice, whereas undercut aversion records an attempt to prevent the other team from obtaining the bonus. See \ref{appendix:ra_instruction} for a detailed description.} The two bounds thus delimit the range of levels consistent with each message, with Stage~1 providing a conservative interpretation and Stage 2 allowing a more generous interpretation. 

In each stage, the two RAs assigned to a class of games first complete their classifications independently and then meet to reconcile any disagreements. We report the reconciled classifications throughout and allow RAs to leave a classification field blank when the message provides insufficient evidence for inference or when the two RAs fail to reach agreement. 
Full classification instructions are provided in \ref{appendix:ra_instruction}.

\section{Experimental Results}

\subsection{The 2/3 Beauty Contest Game}
\label{subsec:beauty_contest_result}

\begin{figure}[htbp!]
    \centering
    \includegraphics[width=\linewidth]{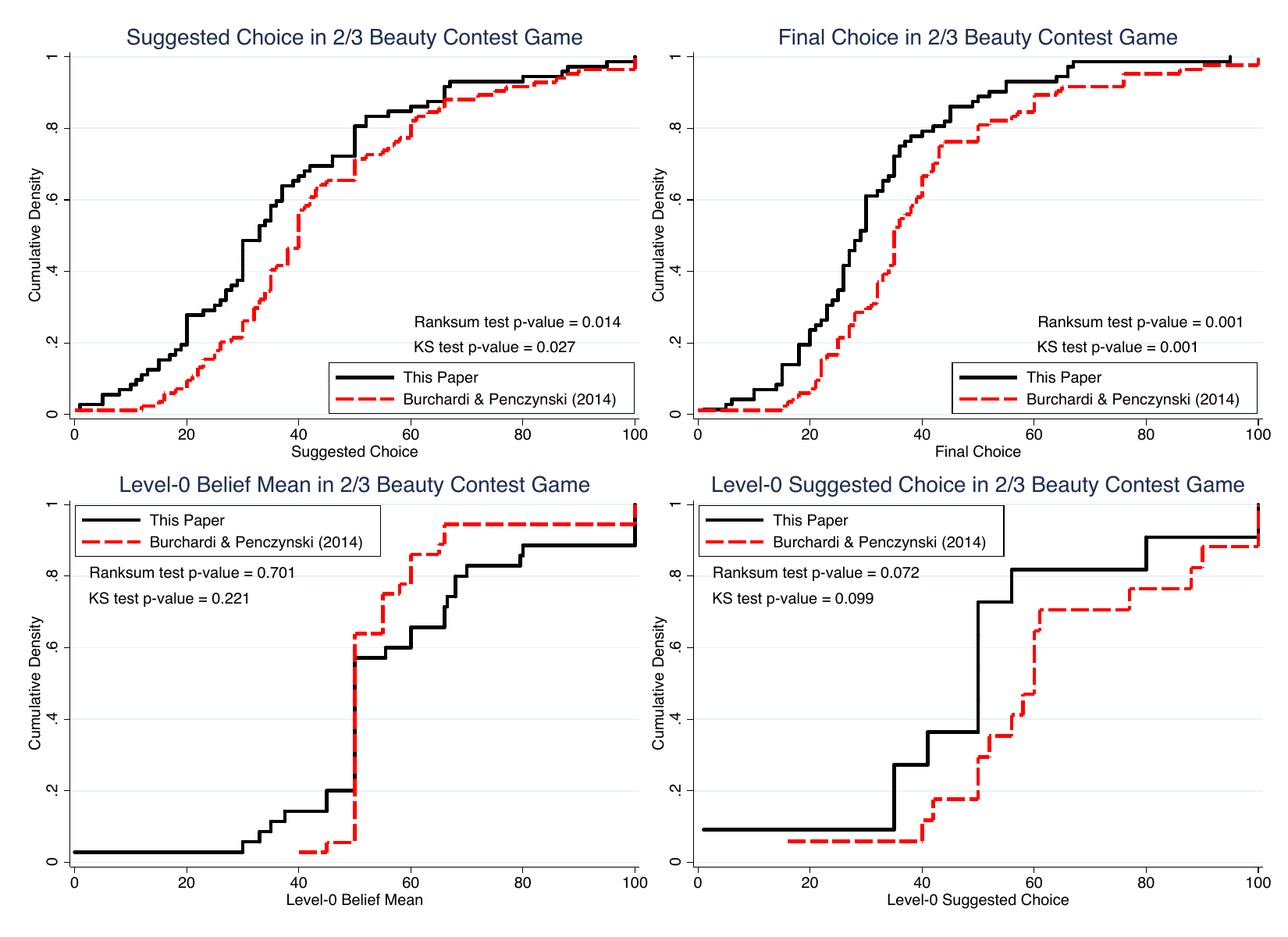}
    \caption{Distributions of Choices and Beliefs in the 2/3 Beauty Contest Game and Comparison with BP. Solid and dashed lines represent our experiment and BP, respectively. (Top left) Suggested choices. (Top right) Final choices. (Bottom left) Level-0 belief means. (Bottom right) Suggested choices among level-0 players.}
    \label{fig:beauty_main_fig}
\end{figure}

Figure~\ref{fig:beauty_main_fig} reports the cumulative distributions of suggested choices, final choices, level-0 belief means, and level-0 suggested choices in the 2/3 beauty contest game for our sample and BP's. The top-left panel shows that our suggested choices are \emph{first-order stochastically dominated} by BP's, with a mean suggested choice of 36.56 in our sample compared with 43.93 in BP (Rank-sum test p-value $= 0.014$; Kolmogorov-Smirnov test p-value $= 0.027$). Furthermore, the same dominance relationship holds for final choices, which are also systematically lower in our sample (Rank-sum test p-value $= 0.001$; Kolmogorov-Smirnov test p-value $= 0.001$). It is worth remarking that our suggested choices are statistically indistinguishable from the choices in the first-round data of \cite{nagel1995unraveling}.\footnote{The mean choice in \cite{nagel1995unraveling}'s first-round $p=2/3$ game is 36.73, close to our 36.56, and the two distributions are statistically indistinguishable (Rank-sum test p-value $ = 0.978$; Kolmogorov-Smirnov test p-value $ = 0.701$).} Thus, we replicate \cite{nagel1995unraveling}'s classic finding in this game, suggesting that the differences between our choice data and BP's are likely attributable to differences in subject pools.


Despite the differences in suggested and final choices, subjects in the two studies hold similar beliefs about the average behavior of non-strategic players. Of our 72 subjects, 35 report a level-0 belief mean, and its distribution is statistically indistinguishable from that of BP's 36 subjects (Rank-sum test p-value $= 0.701$; Kolmogorov-Smirnov test p-value $= 0.221$), as shown in the bottom-left panel of Figure~\ref{fig:beauty_main_fig}. The mean belief is 57.57 in our data and 55.28 in BP, both significantly above the midpoint of 50 (one-sample t-test p-value $= 0.043$ and $0.015$, respectively), consistent with BP's hypothesis that the salience of $p = 2/3$ anchors reasoning above the uniform benchmark.

Nevertheless, a closer look at the composition of these beliefs reveals a modest difference. Thirteen of our 35 beliefs (37\%) are exactly 50, compared with 21 of 36 (58\%) in BP, a marginally significant difference in these proportions (Rank-sum test p-value $= 0.076$). Besides, our belief distribution is also significantly more dispersed than BP's (standard deviation of 21.31 in our experiment versus 12.33 in BP; Levene's test p-value $= 0.016$). Taken together,
these findings indicate that while the two samples are centered on a similar level-0 belief, individual subjects in our sample vary more widely around it.




Moreover, non-strategic players do exist: we identify eleven of our 72 subjects (15\%) as uniquely level-0, with both a lower and an upper bound of 0.\footnote{See \ref{appendix:tables_figures} for the joint distribution of lower and upper bounds.} Their suggested choices average 49.82, compared with 62.35 among BP's 17 level-0 subjects, and our distribution again lies to the left of BP's under first-order stochastic dominance, as plotted in the bottom-right panel of Figure~\ref{fig:beauty_main_fig}. However, the dominance relationship is only marginally significant (Rank-sum test p-value $ = 0.072$; Kolmogorov-Smirnov test p-value $= 0.099$). In addition, the distribution of level-0 suggested choices in our data is widely dispersed, with a standard deviation of 25.23, and is not significantly different from a uniform distribution (one-sample Kolmogorov-Smirnov test p-value $ = 0.332$). By contrast, BP reject uniformity for their level-0 suggested choices using the same test (p-value $ = 0.021$), implying that non-strategic play in our data more closely approximates the uniform benchmark typically assumed for level-0 behavior in the literature.

Lastly, we find that our elicited level-0 belief means are consistent with the suggested choices of our level-0 players. The two distributions, as well as their means, are statistically indistinguishable (Rank-sum test p-value $= 0.278$; Kolmogorov-Smirnov test p-value $ = 0.713$). The same level-0 belief-action consistency also holds in BP's data, suggesting that our replication recovers not only BP's level-0 belief distribution but also the alignment between those beliefs and level-0 actions documented in their study.\footnote{As a side note, three of our subjects identify the equilibrium, while eight apply iterated dominance, compared with two subjects in each category in BP.} Together, the close replication of BP’s findings on level-0 beliefs and actions confirms that the intra-team communication method reliably recovers the reasoning underlying observed choices in the beauty contest game. This recovery gives us confidence to apply the intra-team communication method to the Extreme 11--20 game, despite GLZ's documentation that behavior in this game is starkly inconsistent with the typical findings in level-$k$ literature.

\subsection{The Extreme 11--20 Game}
\label{subsec:11_20_result}

\begin{figure}[htbp!]
    \centering
    \includegraphics[width=\linewidth]{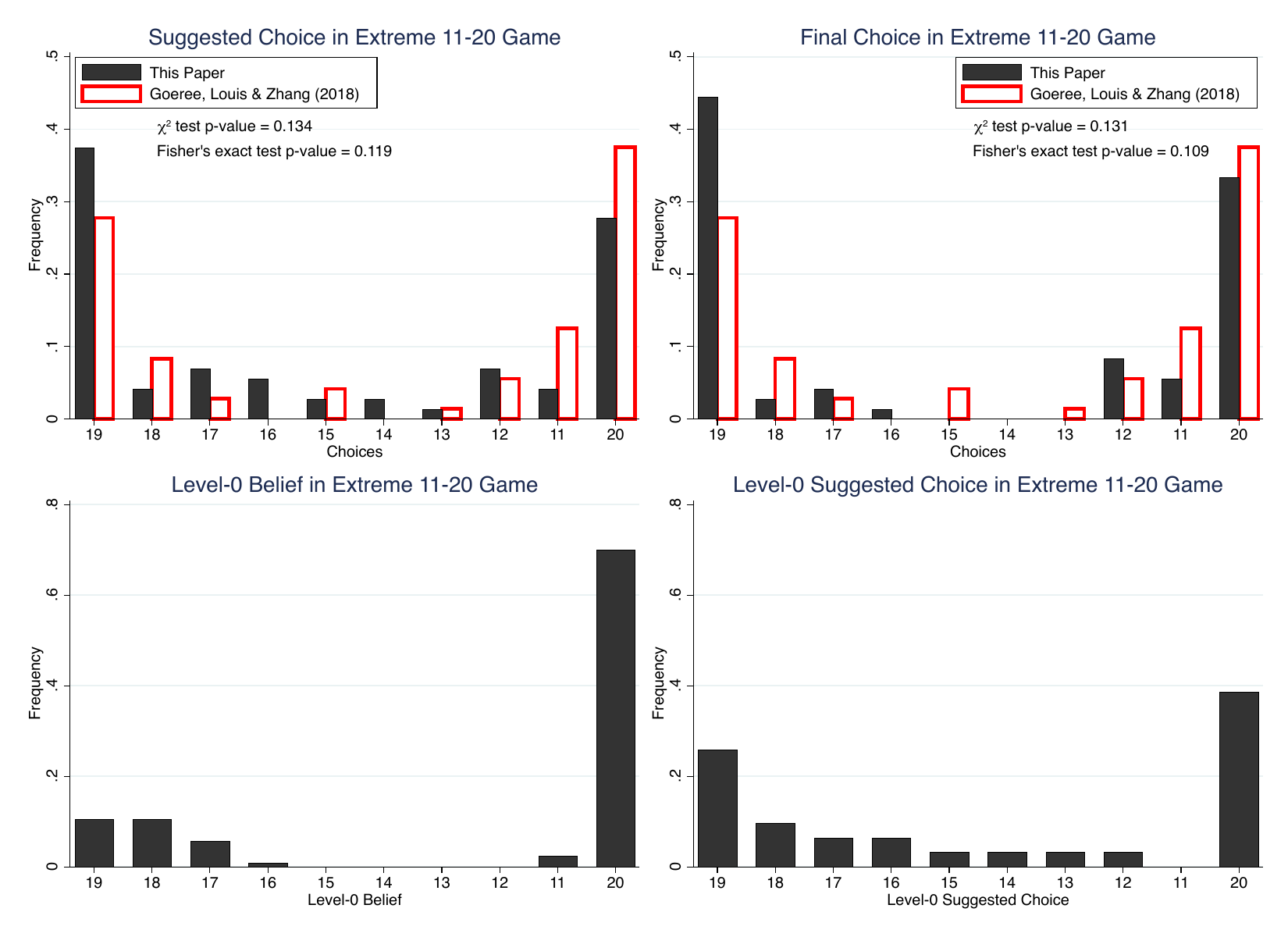}
    \caption{Distribution of Choices and Beliefs in the Extreme 11--20 Game and Comparison with GLZ. Solid bars represent our experiment, and hollow bars represent individual choices from Stage 1 of GLZ. (Top left) Suggested choices. (Top right) Final choices. (Bottom left) Level-0 beliefs. (Bottom right) Suggested choices among level-0 players. The bottom panels display our data only, as GLZ do not elicit level-0 beliefs or actions.}
    \label{fig:main_1120_figure}
\end{figure}

Having validated the intra-team communication method in the beauty contest game, we now apply it to the Extreme 11--20 game, where GLZ document behavior that appears anomalous from the perspective of the standard level-$k$ model. Figure~\ref{fig:main_1120_figure} reports the distributions of suggested choices, final choices, level-0 beliefs, and level-0 suggested choices in our data, alongside GLZ's individual choices.\footnote{
In GLZ, subjects play three versions of the 11--20 game across three stages that differ in their payment schemes. Although there is no feedback across games, for a fair comparison, we focus on their Stage 1 data, in which subjects are first exposed to the Extreme 11--20 game and are paid based on their play against a single randomly selected opponent, as in our design.} Both our suggested and final choices are statistically indistinguishable from those in GLZ (Suggested choices: chi-square test p-value $ = 0.134$ and Fisher's exact test p-value $= 0.119$; Final choices: chi-square test p-value $ = 0.131$ and Fisher's exact test p-value $ = 0.109$), providing the first replication of their central finding that subjects concentrate their choices on the two end boxes, 19 and 20. 
This pattern contrasts sharply with the typical findings in the level-$k$ literature, particularly in the standard 11--20 game (i.e., GLZ's Basic version), where most subjects exhibit one to three steps of reasoning and concentrate their choices on the second, third, and fourth boxes from the right.\footnote{In the Basic 11--20 game, both \cite{arad201211} and GLZ find that a vast majority of subjects choose the numbers corresponding to one to three steps of reasoning (19, 18, and 17, respectively), with proportions of 12\%, 30\%, and 32\% in \cite{arad201211} and 28\%, 40\%, and 11\% in GLZ's first round.}
As a conceptual extension of the replication exercise in Section~\ref{subsec:beauty_contest_result}, we next conduct parallel analyses of level-0 beliefs and actions in the Extreme 11--20 game and explore how these results may help interpret this seemingly anomalous behavior. 

We begin by examining the choices of subjects we classify as level-0.
The bottom-right panel of Figure~\ref{fig:main_1120_figure} shows that our level-0 subjects' suggested choices are far from degenerate at 20: only 12 of 31 (39\%) choose 20, while the remaining 19 are spread across the other nine boxes, with eight choosing 19 and the others scattered from 18 down to 12. This dispersion is inconsistent with the point mass at 20 that the standard level-$k$ model assumes for non-strategic play (one-sample Kolmogorov-Smirnov test p-value $< 0.001$). 


Next, we turn to the level-0 beliefs identified from our message data.
The bottom-left panel of Figure~\ref{fig:main_1120_figure} shows that level-0 beliefs, like level-0 actions, depart significantly from the degenerate belief at 20 assumed in the literature (one-sample Wilcoxon signed-rank test p-value $= 0.005$): 13 of 21 stated beliefs are degenerate at 20, and, together with four mixed beliefs that include 20, place 70\% of the weight on that box, with the remaining 30\% spread mostly across 19, 18, and 17. This departure moves in the same direction as the dispersion we document in level-0's actual choices---away from the point mass at 20---though the belief shifts far less: level-0 subjects place only 39\% of their choices on 20, compared with 70\% of the belief weight higher-level players assign to it. 

We test directly whether level-0 beliefs and level-0 actions differ, accounting for the ambiguity introduced by beliefs that name multiple integers through simulation. In each of 100,000 replications, we draw 21 observations from the empirical belief distribution described above and compare them with the observed level-0 action distribution using a rank-sum test. The median simulated p-value across replications is 0.043, indicating that level-0 beliefs and level-0 actions differ significantly.\footnote{For completeness, we also perform a two-sample Epps–Singleton test on the same simulated draws, which yields a median p-value of 0.013.}
Together, our results show that, unlike in the $2/3$ beauty contest game, level-0 beliefs and actions are misaligned in the Extreme 11--20 game, and both depart from the standard assumption that level-0 action distribution is degenerate at 20.

These results may help make sense of the seemingly anomalous behavior documented by GLZ, namely, the concentration of choices on 19 and 20.
First, around 26\% of level-0 suggested choices are 19, indicating that some observed choices of 19 come from level-0 players themselves.
Moreover, if we use the empirical distribution of level-0 suggested choices as the level-0 specification, 19 becomes the best response to this empirical distribution, even though 20 remains the modal level-0 choice. 
Accordingly, choices of 19 and 20 correspond to levels 1 and 2, respectively, bringing the observed pattern much closer to the reasoning depths typically found in the level-$k$ literature.
Thus, the interpretation of 19 as an implausibly high-level choice is sensitive to the specification of level-0 behavior.
The elicited level-0 beliefs also qualitatively point in this direction.
Because these beliefs are not degenerate at 20 and place some weight on 17 through 19, choices of 19 and 20 can, depending on the level-0 anchor, be consistent with reasoning below level 3.
Overall, our results on level-0 beliefs and actions suggest that GLZ's finding in the Extreme 11--20 game may be less anomalous from the perspective of the standard level-$k$ model under a more empirically grounded specification of level-0 behavior, highlighting the value of extending the intra-team communication method to another canonical class of games in the level-$k$ literature.\footnote{Beyond the level-$k$ framework, the Extreme version may also induce security concerns about one's own payoff or being undercut, which could help account for choices at both ends. Among the 47 subjects choosing 19 or 20, 34 (72\%) invoke payoff security, 7 (15\%) undercut aversion, and 4 (9\%) both.}

\section{Discussion and Concluding Remarks}
A key advantage of the intra-team communication method is that it provides process data on individual strategic reasoning, allowing researchers to identify subjects’ levels of reasoning beyond what can be inferred from observed choices alone.
To illustrate this advantage further, we leverage our within-subject design to explore the cross-game stability of individual reasoning depth, an important question in the level-$k$ literature.

Recent studies have shown that experimental subjects exhibit limited stability in reasoning depth across different games when interacting with human opponents \citep{georganas2015persistence, cooper2024consistent, chen2025measuring}, raising the possibility that individual reasoning depth is context-dependent rather than a persistent personal trait.
In these experiments, however, an individual's level of reasoning is estimated using choice data alone.
An alternative hypothesis is therefore that some of the observed instability may reflect noise in level-$k$ estimation arising from misspecified heterogeneous level-0 behavior and beliefs.\footnote{Another possibility is that subjects hold heterogeneous beliefs about opponents' sophistication: \cite{chen2025measuring} find greater cross-game stability in reasoning levels against robot players of known levels.}

\begin{figure}[htbp!]
    \centering
    \includegraphics[width=0.8\linewidth]{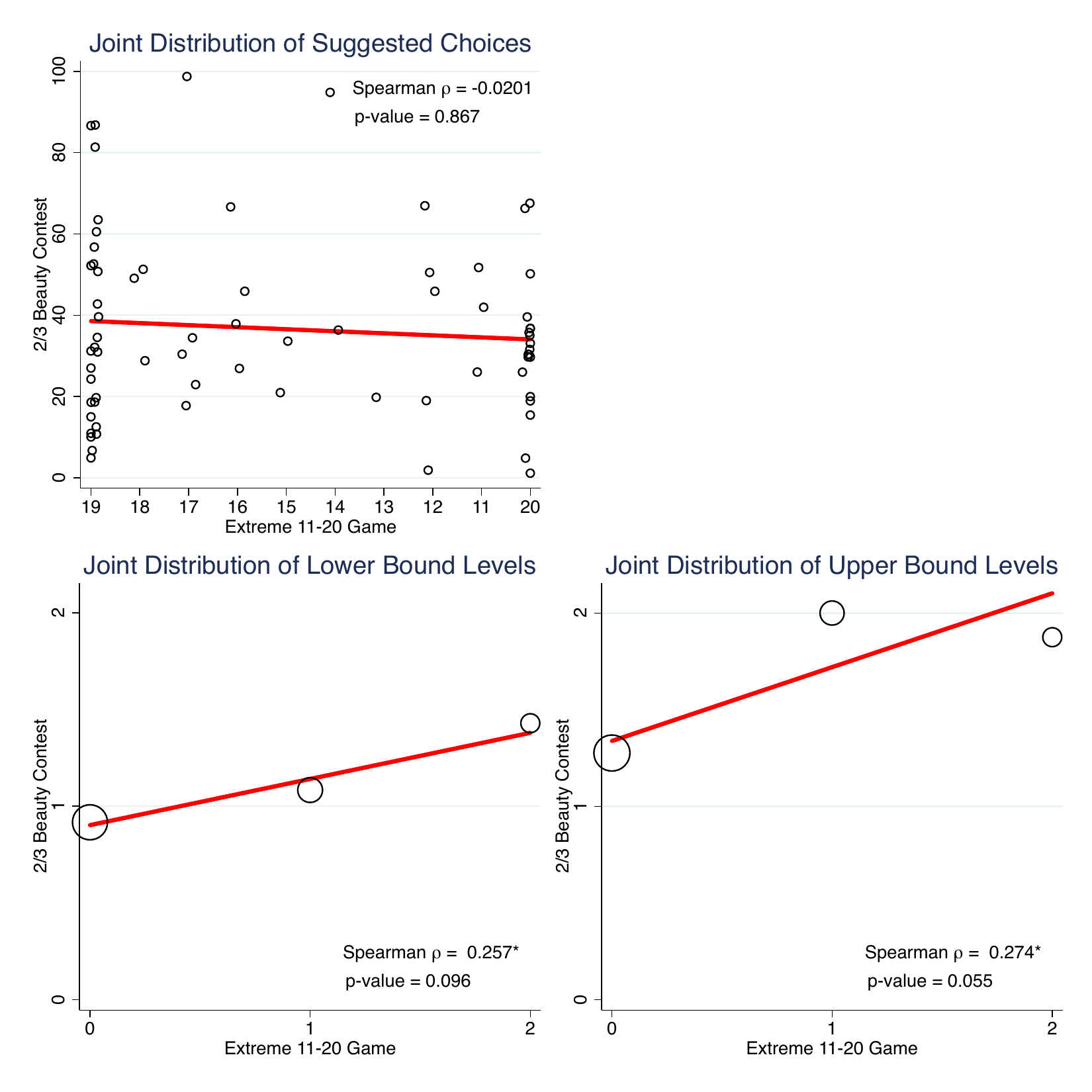}
    \caption{Correlation of Choices and Levels across the 2/3-Beauty Contest and the Extreme 11--20 Game. (Top) Suggested choices. (Bottom left) Reasoning levels from lower-bound classifications. (Bottom right) Reasoning levels from upper-bound classifications. In the bottom panels, the height of each circle represents the mean level in the 2/3-beauty contest among subjects assigned a given level in the Extreme 11--20, and the size of the circle is proportional to the number of such subjects. Regression lines are shown in all panels.}
    \label{fig:correlation_fig}
\end{figure}

Using reasoning levels inferred directly from messages in our beauty contest and 11--20 games, we find evidence consistent with this alternative hypothesis. 
Figure~\ref{fig:correlation_fig} reports the joint distributions across the two games of suggested choices (top panel) and reasoning levels based on the lower-bound (bottom-left panel) and upper-bound (bottom-right panel) classifications.
Under the standard level-$k$ model, smaller numbers in the beauty contest game and positions farther to the left in the 11--20 game correspond to higher reasoning levels.
Using choices as proxies for reasoning depth, we find essentially no correlation across the two games (Spearman $\rho=-0.020$, p-value $=0.867$).

In contrast, among subjects whose reasoning levels are identified from messages in both games, we observe a positive cross-game correlation under both the lower-bound classification (Spearman $\rho=0.257$, p-value $=0.096$) and the upper-bound classification (Spearman $\rho=0.274$, p-value $=0.055$).
In other words, even when choices appear uncorrelated across different types of games, incentivized message data reveal some cross-game stability in reasoning depth without requiring ad hoc assumptions about level-0 behavior.
This finding qualifies previous evidence of instability in level-$k$ reasoning and highlights a further advantage of the intra-team communication method.

To summarize, we conduct an experiment that replicates BP’s beauty contest game and incorporates their intra-team communication method into GLZ’s variants of the 11--20 game, allowing us to identify subjects’ reasoning levels and elicit their beliefs about non-strategic play using incentivized message data.
In the $2/3$-beauty contest game, we closely replicate BP’s findings on level-0 beliefs and their consistency with observed level-0 behavior. 
This consistency breaks down in the Extreme 11--20 game, where elicited beliefs and observed level-0 actions diverge. 
Moreover, both level-0 beliefs and actions depart from the standard assumption that level-0 players choose 20 with certainty, offering a complement to GLZ's noisy-introspection explanation for the observed behavioral pattern.
Finally, our exploratory cross-game analysis suggests that some apparent instability in reasoning depth documented in the literature may reflect the limitations of inferring levels from choices alone, illustrating an additional methodological value of the intra-team communication protocol beyond its existing applications.

\bibliographystyle{ecta}
\bibliography{reference}

\newpage
\appendix

\setcounter{section}{0}
\renewcommand{\thesection}{Appendix \Alph{section}}
\setcounter{figure}{0}
\renewcommand{\thefigure}{A\arabic{figure}}
\setcounter{table}{0}
\renewcommand{\thetable}{A\arabic{table}}

\section{Additional Tables and Figures}
\label{appendix:tables_figures}

\subsection*{Suggested Choices, Final Choices, and Level-0 Behavior}

\begin{figure}[H]
    \centering
    \includegraphics[width=\linewidth]{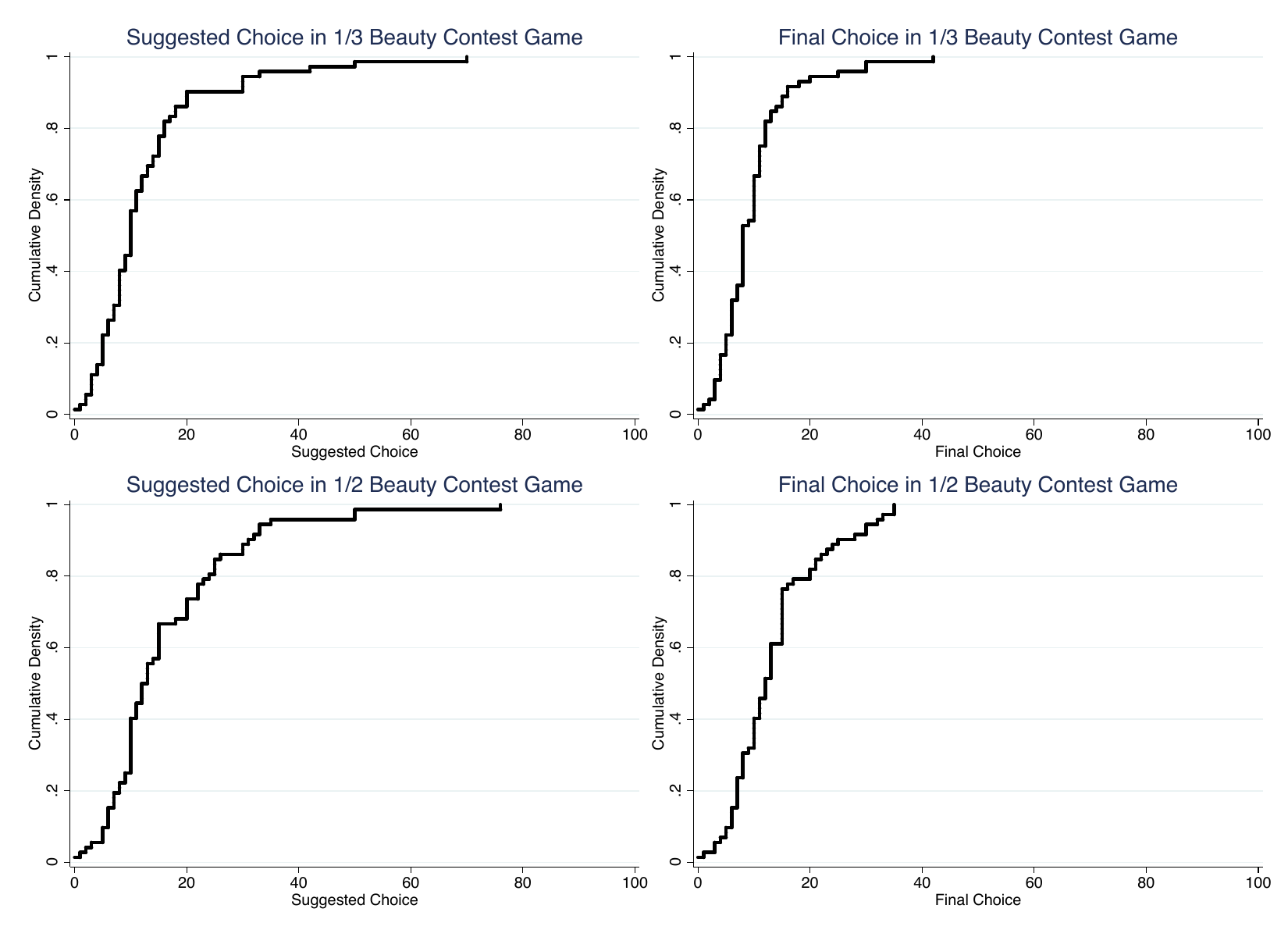}
    \caption{Distributions of Suggested and Final Choices in the 1/3 and 1/2 Beauty Contest Games. (Top row) 1/3 beauty contest game. (Bottom row) 1/2 beauty contest game.}
    \label{fig:appendix_beauty_choice_figure}
\end{figure}

\begin{figure}[H]
    \centering
    \includegraphics[width=\linewidth]{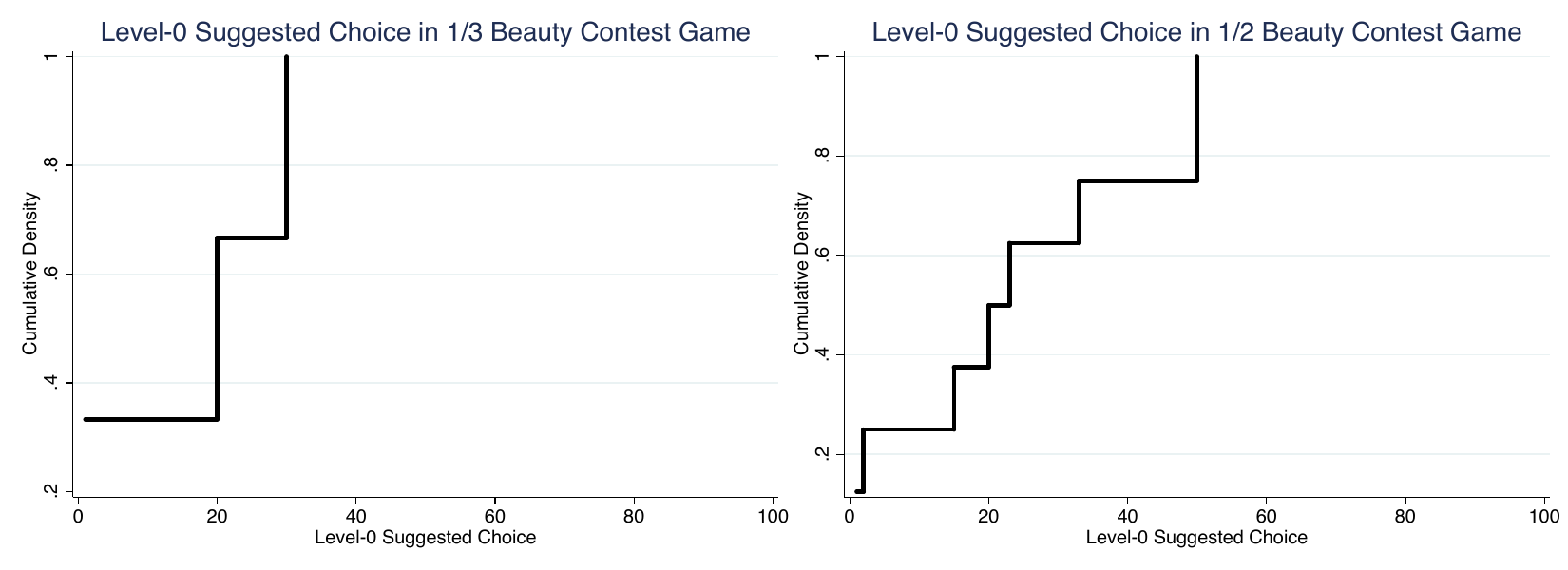}
    \caption{Distributions of Suggested Choices among Level-0 Players in the 1/3 and 1/2 Beauty Contest Games. (Left) 1/3 beauty contest game. (Right) 1/2 beauty contest game.}
    \label{fig:appendix_beauty_level_0_choice_figure}
\end{figure}

\begin{figure}[H]
    \centering
    \includegraphics[width=\linewidth]{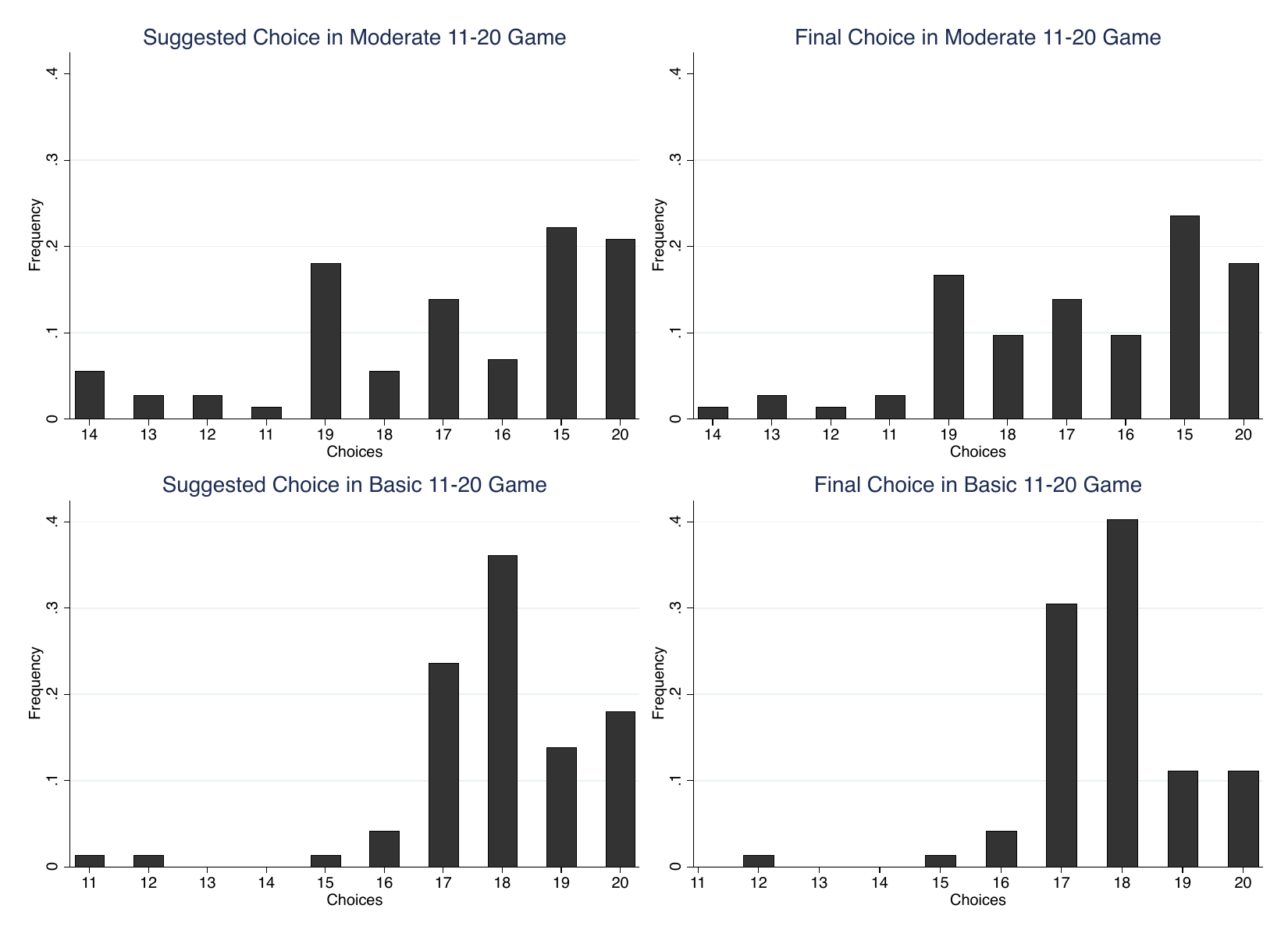}
    \caption{Distributions of Suggested and Final Choices in the Moderate and Basic 11--20 Games. (Top row) Moderate 11--20 game. (Bottom row) Basic 11--20 game.}
    \label{fig:appendix_1120_choice_figure}
\end{figure}

\begin{figure}[H]
    \centering
    \includegraphics[width=\linewidth]{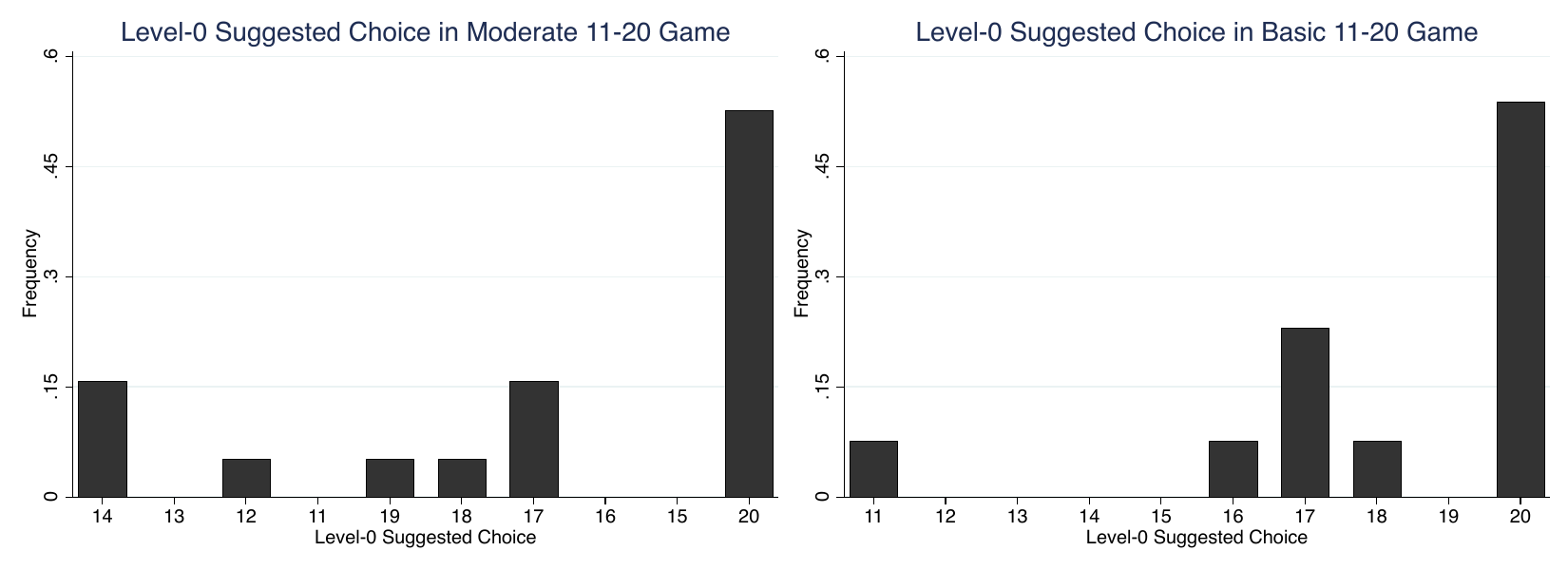}
    \caption{Distributions of Suggested Choices among Level-0 Players in the Moderate and Basic 11--20 Games. (Left) Moderate 11--20 game. (Right) Basic 11--20 game.}
    \label{fig:appendix_1120_level_0_choice_figure}
\end{figure}

Tables~\ref{tab:summary_beauty} and \ref{tab:11--20} report summary statistics for suggested choices, final choices, and level-0 suggested choices across all six games, complementing the distributions for the first round of each block shown in Figures~\ref{fig:beauty_main_fig} and \ref{fig:main_1120_figure}. Figures~\ref{fig:appendix_beauty_choice_figure}-\ref{fig:appendix_1120_level_0_choice_figure} plot the corresponding distributions for the remaining four games: Figures~\ref{fig:appendix_beauty_choice_figure} and \ref{fig:appendix_beauty_level_0_choice_figure} cover the 1/3 and 1/2 beauty contest games, and Figures~\ref{fig:appendix_1120_choice_figure} and \ref{fig:appendix_1120_level_0_choice_figure} cover the Moderate and Basic 11--20 games. Because the box order differs across the three versions of the 11--20 game, Table~\ref{tab:11--20} reports summary statistics for the \emph{locations} of the choices, with the leftmost box coded as 1 and the rightmost box as 10, rather than by the numerical labels displayed in the boxes.

\begin{table}[H]
\centering
\caption{Summary Statistics for the Beauty Contest Games}
\label{tab:summary_beauty}
\renewcommand{\arraystretch}{1.25}
\begin{tabular}{lcccccccc}
\hline
 & Game &  & Obs. & Mean & S.D. & Median & Max & Min \\ \hline
\multicolumn{9}{l}{\emph{Panel A. Suggested Choice}} \\
 & $p=2/3$ &  & 72 & 36.56 & 21.91 & 33 & 100 & 1 \\
 & $p=1/3$ &  & 72 & 12.60 & 11.29 & 10 & 70 & 0 \\
 & $p=1/2$ &  & 72 & 16.31 & 12.45 & 12.5 & 76 & 0 \\ \hline
\multicolumn{9}{l}{\emph{Panel B. Final Choice}} \\
 & $p=2/3$ &  & 72 & 31.36 & 16.15 & 29 & 95 & 1 \\
 & $p=1/3$ &  & 72 & 9.81 & 6.82 & 8 & 42 & 0 \\
 & $p=1/2$ &  & 72 & 13.56 & 8.05 & 12 & 35 & 0 \\ \hline
\multicolumn{9}{l}{\emph{Panel C. Level-0 Suggested Choice}} \\
 & $p=2/3$ &  & 11 & 49.82 & 25.23 & 50 & 100 & 1 \\
 & $p=1/3$ &  & 3 & 17.00 & 14.73 & 20 & 30 & 1 \\
 & $p=1/2$ &  & 8 & 24.25 & 19.06 & 21.5 & 50 & 1 \\ \hline
\end{tabular}
\end{table}

\begin{table}[H]
\centering
\caption{Summary Statistics for the 11--20 Games}
\label{tab:11--20}
\renewcommand{\arraystretch}{1.25}
\begin{threeparttable}
\begin{tabular}{lcccccccc}
\hline
 & Game &  & Obs. & Mean & S.D. & Median & Max & Min \\ \hline
\multicolumn{9}{l}{\emph{Panel A. Suggested Choice (Location)}} \\
 & Extreme &  & 72 & 5.00 & 3.90 & 4 & 10 & 1 \\
 & Moderate &  & 72 & 7.10 & 2.64 & 7.5 & 10 & 1 \\
 & Basic &  & 72 & 7.96 & 1.61 & 8 & 10 & 1 \\ \hline
\multicolumn{9}{l}{\emph{Panel B. Final Choice (Location)}} \\
 & Extreme &  & 72 & 5.18 & 4.21 & 3 & 10 & 1 \\
 & Moderate &  & 72 & 7.32 & 2.25 & 8 & 10 & 1 \\
 & Basic &  & 72 & 7.82 & 1.28 & 8 & 10 & 2 \\ \hline
\multicolumn{9}{l}{\emph{Panel C. Level-0 Suggested Choice (Location)}} \\
 & Extreme &  & 31 & 5.61 & 3.95 & 5 & 10 & 1 \\
 & Moderate &  & 19 & 7.26 & 3.48 & 10 & 10 & 1 \\
 & Basic &  & 13 & 8.15 & 2.64 & 10 & 10 & 1 \\ \hline
\end{tabular}
\begin{tablenotes}
\item {\footnotesize Since the order of the boxes differs across the three versions of the game, we report summary statistics for the \emph{locations} where the leftmost box is coded as 1 and the rightmost box as 10, rather than for the actual numbers shown in the boxes.}
\end{tablenotes}
\end{threeparttable}
\end{table}

\subsection*{Level-0 Beliefs}

\begin{figure}[H]
    \centering
    \includegraphics[width=\linewidth]{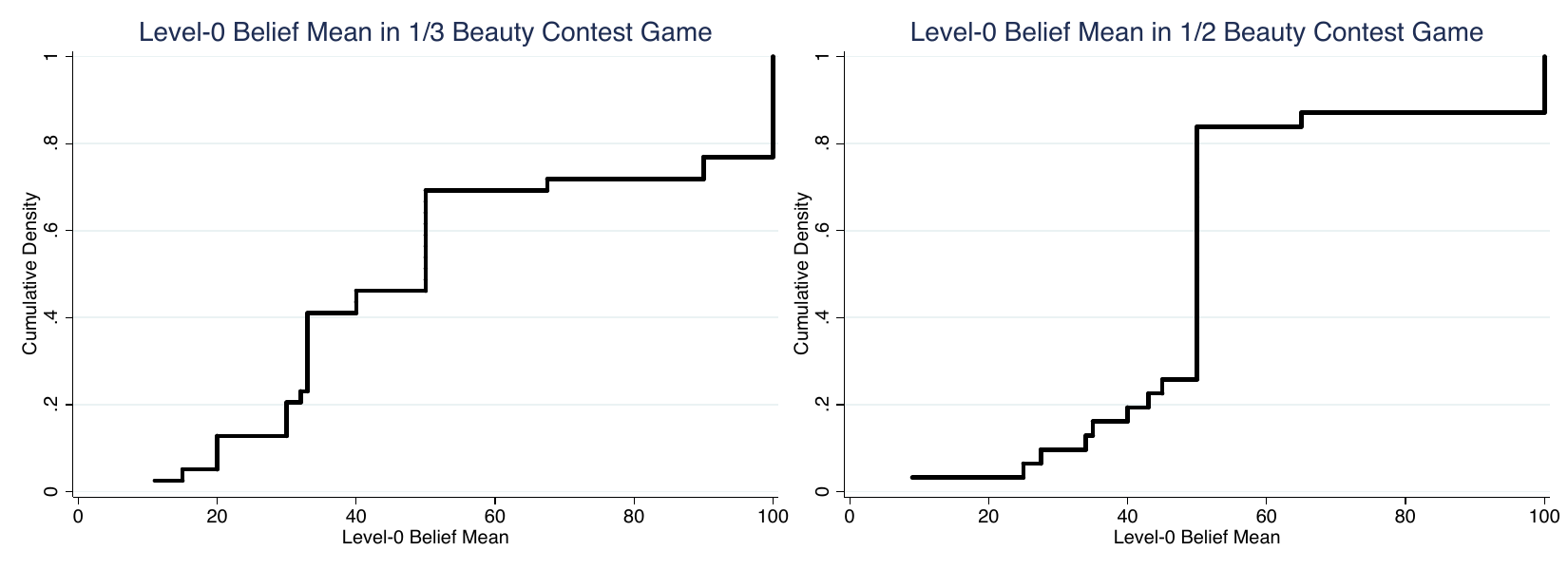}
        \caption{Distributions of Level-0 Belief Means in the 1/3 and 1/2 Beauty Contest Games. (Left) 1/3 beauty contest game. (Right) 1/2 beauty contest game.}
    \label{fig:appendix_beauty_level_0_belief_figure}
\end{figure}

Table \ref{tab:summary_beauty_belief} and Figure \ref{fig:appendix_beauty_level_0_belief_figure} report summary statistics and distributions of level-0 belief means in the three beauty contest games, and Table \ref{tab:summary_1120_belief} reports the frequency of level-0 beliefs by box location in the three 11--20 games.

\begin{table}[H]
\centering
\caption{Summary Statistics for Level-0 Beliefs in the Beauty Contest Games}
\label{tab:summary_beauty_belief}
\renewcommand{\arraystretch}{1.25}
\begin{tabular}{lcccccccc}
\hline
 & Game &  & Obs. & Mean & S.D. & Median & Max & Min \\ \hline
 & $p=2/3$ &  & 35 & 57.57 & 21.31 & 50 & 100 & 0 \\
 & $p=1/3$ &  & 39 & 54.27 & 30.13 & 50 & 100 & 11 \\
 & $p=1/2$ &  & 31 & 52.37 & 21.23 & 50 & 100 & 9 \\ \hline
\end{tabular}
\end{table}

\begin{table}[H]
\centering
\caption{Frequencies of Level-0 Beliefs in the 11--20 Games}
\label{tab:summary_1120_belief}
\renewcommand{\arraystretch}{1.3}
\begin{adjustbox}{width=\columnwidth,center}
\begin{threeparttable}
\begin{tabular}{@{}ccccccccccccc@{}}
\toprule
 &  &  & \multicolumn{10}{c}{Box Location} \\ \cmidrule(l){4-13} 
Game & Obs. &  & 1 & 2 & 3 & 4 & 5 & 6 & 7 & 8 & 9 & 10 \\ \midrule
Extreme & 21 &  & 0.105 & 0.105 & 0.057 & 0.010 & 0 & 0 & 0 & 0 & 0.024 & 0.700 \\
Moderate & 48 &  & 0.021 & 0 & 0 & 0 & 0.186 & 0.061 & 0.010 & 0.010 & 0.016 & 0.696 \\
Basic & 44 &  & 0 & 0 & 0 & 0 & 0 & 0 & 0.011 & 0.045 & 0.045 & 0.898 \\ \bottomrule
\end{tabular}
\begin{tablenotes}
\item {\footnotesize Since the order of the boxes differs across the three versions of the game, we report the frequencies for the \emph{locations}, where the leftmost box is coded as 1 and the rightmost box as 10, rather than for the actual numbers shown in the boxes. If multiple integers are specified in a message, we assume a uniform distribution over that support.}
\end{tablenotes}
\end{threeparttable}
\end{adjustbox}
\end{table}

\subsection*{Level Classification Results}

Tables \ref{tab:classification_table_2_3_beauty}–\ref{tab:classification_table_1_2_beauty} report the joint distribution of lower and upper bounds for the three beauty contest games, and Tables \ref{tab:classification_table_extreme_1120}–\ref{tab:classification_table_basic_1120} report the same for the three 11--20 games, following BP's two-stage classification approach described in Section \ref{subsec:message_classification}. In each table, rows correspond to the lower bound recorded in Stage 1 and columns to the upper bound recorded in Stage 2, and each cell reports the number of subjects classified with that combination of bounds. An entry of N/A indicates that no bound could be assigned to the corresponding message, either for lack of evidence or because the two RAs could not reach agreement.

\begin{table}[H]
\centering
\caption{Level Classification Results for the 2/3 Beauty Contest Game}
\label{tab:classification_table_2_3_beauty}
\renewcommand{\arraystretch}{1.25}
\normalsize
\begin{tabular}{cc|*{6}{>{\centering\arraybackslash}p{1cm}}|c}
\hline
 &  & \multicolumn{6}{c|}{Level Upper Bound} & Total \\ \cline{3-8}
 &  & 0 & 1 & 2 & 3 & 4 & N/A &  \\ \hline
 & 0 & 11 & 0 & 0 & 0 & 0 & 2 & 13 \\
 & 1 & 1 & 10 & 16 & 2 & 0 & 2 & 31 \\
Level Lower & 2 & 0 & 0 & 7 & 6 & 3 & 0 & 16 \\
Bound & 3 & 0 & 0 & 0 & 0 & 0 & 0 & 0 \\
 & 4 & 0 & 0 & 0 & 0 & 0 & 0 & 0 \\
 & N/A & 3 & 0 & 4 & 1 & 0 & 4 & 12 \\ \hline
Total &  & 15 & 10 & 27 & 9 & 3 & 8 & 72 \\ \hline
\end{tabular}
\end{table}

\begin{table}[H]
\centering
\caption{Level Classification Results for the 1/3 Beauty Contest Game}
\label{tab:classification_table_1_3_beauty}
\renewcommand{\arraystretch}{1.25}
\normalsize
\begin{tabular}{cc|*{6}{>{\centering\arraybackslash}p{1cm}}|c}
\hline
 &  & \multicolumn{6}{c|}{Level Upper Bound} & Total \\ \cline{3-8}
 &  & 0 & 1 & 2 & 3 & 4 & N/A &  \\ \hline
 & 0 & 3 & 1 & 1 & 0 & 0 & 2 & 7 \\
 & 1 & 0 & 2 & 14 & 5 & 0 & 1 & 22 \\
Level Lower & 2 & 0 & 0 & 10 & 9 & 3 & 0 & 22 \\
Bound & 3 & 0 & 0 & 0 & 1 & 1 & 0 & 2 \\
 & 4 & 0 & 0 & 0 & 0 & 0 & 0 & 0 \\
 & N/A & 0 & 0 & 14 & 2 & 0 & 3 & 19 \\ \hline
Total &  & 3 & 3 & 39 & 17 & 4 & 6 & 72 \\ \hline
\end{tabular}
\end{table}

\begin{table}[H]
\centering
\caption{Level Classification Results for the 1/2 Beauty Contest Game}
\label{tab:classification_table_1_2_beauty}
\renewcommand{\arraystretch}{1.25}
\normalsize
\begin{tabular}{cc|*{6}{>{\centering\arraybackslash}p{1cm}}|c}
\hline
 &  & \multicolumn{6}{c|}{Level Upper Bound} & Total \\ \cline{3-8}
 &  & 0 & 1 & 2 & 3 & 4 & N/A &  \\ \hline
 & 0 & 8 & 2 & 0 & 0 & 0 & 3 & 13 \\
 & 1 & 0 & 2 & 6 & 3 & 0 & 0 & 11 \\
Level Lower & 2 & 0 & 0 & 12 & 13 & 1 & 1 & 27 \\
Bound & 3 & 0 & 0 & 0 & 3 & 0 & 0 & 3 \\
 & 4 & 0 & 0 & 0 & 0 & 0 & 0 & 0 \\
 & N/A & 1 & 0 & 5 & 0 & 0 & 12 & 18 \\ \hline
Total &  & 9 & 4 & 23 & 19 & 1 & 16 & 72 \\ \hline
\end{tabular}
\end{table}

\begin{table}[H]
\centering
\caption{Level Classification Results for the Extreme 11--20 Game}
\label{tab:classification_table_extreme_1120}
\renewcommand{\arraystretch}{1.25}
\normalsize
\begin{tabular}{cc|*{6}{>{\centering\arraybackslash}p{1cm}}|c}
\hline
 &  & \multicolumn{6}{c|}{Level Upper Bound} & Total \\ \cline{3-8}
 &  & 0 & 1 & 2 & 3 & 4 & N/A &  \\ \hline
 & 0 & 31 & 2 & 0 & 0 & 0 & 1 & 34 \\
 & 1 & 0 & 10 & 2 & 0 & 0 & 1 & 13 \\
Level Lower & 2 & 0 & 0 & 7 & 0 & 0 & 0 & 7 \\
Bound & 3 & 0 & 0 & 0 & 0 & 0 & 0 & 0 \\
 & 4 & 0 & 0 & 0 & 0 & 0 & 0 & 0 \\
 & N/A & 4 & 2 & 0 & 0 & 0 & 12 & 18 \\ \hline
Total &  & 35 & 14 & 9 & 0 & 0 & 14 & 72 \\ \hline
\end{tabular}
\end{table}

\begin{table}[H]
\centering
\caption{Level Classification Results for the Moderate 11--20 Game}
\label{tab:classification_table_moderate_1120}
\renewcommand{\arraystretch}{1.25}
\normalsize
\begin{tabular}{cc|*{6}{>{\centering\arraybackslash}p{1cm}}|c}
\hline
 &  & \multicolumn{6}{c|}{Level Upper Bound} & Total \\ \cline{3-8}
 &  & 0 & 1 & 2 & 3 & 4 & N/A &  \\ \hline
 & 0 & 19 & 0 & 0 & 0 & 0 & 0 & 19 \\
 & 1 & 0 & 20 & 2 & 1 & 0 & 0 & 23 \\
Level Lower & 2 & 0 & 0 & 10 & 2 & 0 & 0 & 12 \\
Bound & 3 & 0 & 0 & 0 & 3 & 0 & 1 & 4 \\
 & 4 & 0 & 0 & 0 & 0 & 0 & 0 & 0 \\
 & N/A & 8 & 6 & 0 & 0 & 0 & 0 & 14 \\ \hline
Total &  & 27 & 26 & 12 & 6 & 0 & 1 & 72 \\ \hline
\end{tabular}
\end{table}

\begin{table}[H]
\centering
\caption{Level Classification Results for the Basic 11--20 Game}
\label{tab:classification_table_basic_1120}
\renewcommand{\arraystretch}{1.25}
\normalsize
\begin{tabular}{cc|*{6}{>{\centering\arraybackslash}p{1cm}}|c}
\hline
 &  & \multicolumn{6}{c|}{Level Upper Bound} & Total \\ \cline{3-8}
 &  & 0 & 1 & 2 & 3 & 4 & N/A &  \\ \hline
 & 0 & 13 & 2 & 0 & 0 & 0 & 1 & 16 \\
 & 1 & 1 & 6 & 7 & 0 & 0 & 1 & 15 \\
Level Lower & 2 & 0 & 0 & 17 & 0 & 0 & 0 & 17 \\
Bound & 3 & 0 & 0 & 0 & 9 & 0 & 0 & 9 \\
 & 4 & 0 & 0 & 0 & 0 & 2 & 0 & 2 \\
 & N/A & 6 & 1 & 2 & 1 & 0 & 3 & 13 \\ \hline
Total &  & 20 & 9 & 26 & 10 & 2 & 5 & 72 \\ \hline
\end{tabular}
\end{table}

\newpage
\section[Experimental Instructions]{Experimental Instructions\footnote{This appendix provides the experimental instructions for the treatment order in which subjects first play three rounds of the beauty contest game, followed by three rounds of the 11--20 game. The instructions for the other treatment order are identical, except that Parts One and Two are swapped.}}
\label{appendix:instructions}

\noindent\textbf{General Instructions}

\medskip

\noindent You are about to participate in an experiment in team decision making. Your earnings will depend partly on your decisions, partly on the decision of others, and partly on chance.  

\bigskip

\noindent Please turn off all your mobile devices. The entire session will take place through computer terminals, and all interactions between participants will be conducted through the computers. Do not open any applications on the computer unless you are instructed to do so. Do not talk or in any way try to communicate with other participants during the session, except according to our instructions. 

\bigskip

\noindent The experiment consists of two parts (\textbf{Part I} and \textbf{Part II}), which are independent of each other and feature different tasks, with three rounds in each part. You will receive the instructions for Part II after Part I has concluded. If you have any questions during the instructions, please raise your hand. The experimenter will approach you and answer your question confidentially. However, if the question appears relevant to others or reflects a common misunderstanding, it will be answered so that everyone can hear, for clarity and transparency.

\bigskip

\noindent Since this is a team experiment, you will at various times be matched randomly with another participant in this room, to form a team that plays as one entity. Your team’s earnings will be shared equally between you and your team partner. The way you interact as a team to take decisions will be the same throughout the two parts.

\bigskip

\noindent Now, let us explain how your \emph{\textbf{Team’s Action}} is determined. In each round, both your team partner and you will enter a \emph{\textbf{Final Decision}} individually and the computer will choose randomly which one of your two final decisions counts as your team’s action. The probability that your team partner’s final decision is chosen is equal to the probability that your final decision will be chosen (i.e. your chances are 50:50). However, you have the possibility to influence your partner’s final decision in the following way: Before you enter your final decision, you can propose to your partner a \emph{\textbf{Suggested Decision}} and send her/him one and only one text \emph{\textbf{Message}}. \emph{Note that this message is your only chance to convince your partner of the reasoning behind your suggested decision. Therefore, use the message to explain your suggested decision to your team partner.} After you finish entering your suggested decision and your message, these will be shown to your team partner. She/he will then make her/his final decision. Similarly, you will receive your partner’s suggested decision and message. You will then make your final decision. As outlined above, once you both enter your final decision, the computer chooses randomly one of your final decisions as your team’s action.

\bigskip

\noindent In this experiment, you will earn ``points'' in each round. Your earnings will be determined by the sum of all the points you earn across both parts. Each point has a value of 25 cents. That is, every 4 points generates \$1 in earnings for you. In addition to your earnings from decisions, you will receive a show-up fee of \$10. At the end of the session, we will record your earnings and you will be paid via a UVA-issued electronic check through PaymentWorks. Our department staff will email you with payment details.

\bigskip

\noindent In order for you to get familiar with the messaging system, you will now try it out in a \textbf{Test Period}. 

\bigskip
\bigskip

\noindent \textbf{Test Period}

\medskip

\noindent A participant in this room is now randomly chosen to be your team partner. The \textbf{Test Period} has two rounds, with one question to answer in each round. Since this is only a test, your earnings will \textbf{not} depend on any decision taken now. In both test rounds you will need to answer a question about the year of a historical event. The team that is closest to the correct year wins.

\bigskip

\noindent As described, you will be able to send one \emph{\textbf{Suggested Decision}} with your proposed year and an explaining \emph{\textbf{Message}}. After having read your partner’s suggested decision and message, you will enter your \emph{\textbf{Final Decision}}. As described earlier, either your or your partner’s final decision will be chosen randomly to be your \emph{\textbf{Team’s Action}}.

\bigskip

\noindent You can send \emph{\textbf{Messages}} of any length by typing in the text box on the screen. You can delete any part of your message using the ``Backspace'' key on your keyboard. Once you have finished typing your message, please click the ``Send Message'' button to send it to your team partner.

\bigskip

\noindent When you are ready, please click the ``Ready'' button to start the Test Period.

\bigskip
\bigskip

\noindent \textbf{Start Part I}

\medskip

\noindent You are about to start \textbf{Part I} of the experiment. You are now randomly matched with a new team partner. For each of the next three rounds you will be matched with a new team partner, i.e. in each of the following rounds you will play with a different person.

\bigskip

\noindent In each round, once all final decisions have been taken, the computer will record which of the final decisions (yours or your team partner’s) is chosen randomly as your team’s action, whether your team won the round, and your personal earnings. \textbf{In each round, the winning team earns 100 points (50 points per team member).} You will be informed about all the records in this and the other part at the end of the experiment, but not between rounds.

\bigskip

\noindent Then the next round of the game follows. It will feature an identical task, but you will be matched with a new team partner.

\bigskip

\noindent Your task is the following: 

\bigskip

\noindent Your team and all other teams will take their Team’s Action by choosing a number between 0 and 100. 0 and 100 are also possible. Only whole numbers will be accepted. From all teams’ actions, the computer will calculate the \emph{Average}. A fraction $p$ of this average will be your \emph{target} number. The winning team will be the one that is \emph{closest to p times the average}. For example, if $p$ is two thirds (2/3), then the winning team will be the one that is closest to two thirds of the average. If two or more teams are equally close, the prize will be randomly given to one of these teams.

\bigskip

\noindent \textbf{Important:} The value of $p$ may change from round to round. At the beginning of each round, the screen will display the value of $p$ for that round. All participants will see the same value of $p$ as you do.

\bigskip

\noindent As described earlier, you will send your team partner a \emph{\textbf{Suggested Decision}} and a \emph{\textbf{Message}}. Remember to explain in the message your reasoning behind your suggested decision. After this information is exchanged, both of you enter your \emph{\textbf{Final Decision}}, from which the computer randomly chooses the \emph{\textbf{Team’s Action}}.

\bigskip

\noindent You will be paid the sum of all points you earn in the three rounds (plus the points you earn in the other part).

\bigskip

\noindent When you click the ``Ready'' button, you will start the first round of \textbf{Part I} of the experiment.

\bigskip
\bigskip

\noindent \textbf{Start Part II}

\medskip

\noindent You are about to start \textbf{Part II} of the experiment. You are now randomly matched with a new team partner. For each of the next three rounds you will be matched with a new team partner, i.e. in each of the following rounds you will play with a different person.

\bigskip

\noindent In each round, once all final decisions have been taken, the computer will record which of the final decisions (yours or your team partner’s) is chosen randomly as your team’s action, how many points your team received, and your personal earnings. You will be informed about all the records in this and the other part at the end of the experiment, but not between rounds.

\bigskip

\noindent Your task is the following: 

\bigskip

\noindent Your team and another team in the experiment are randomly matched to play the following game.
\begin{center}
    \textbf{13. \;  16. \; 20. \; 18. \; 19. \;  12. \;  15. \; 11. \; 17. \; 14}
\end{center}

\noindent On your screen you will see 10 boxes in line, containing different amounts (like the ones above). Each team will take its \emph{Team’s Action} by individually selecting one of the 10 boxes. Each team will receive the amount in the box corresponding to its Team’s Action. \textbf{A team will receive an additional amount of 20 points if its selected amount is exactly ``one to the left'' of the box that the other team chooses.}

\bigskip

\noindent \textbf{Example:} Suppose the boxes are arranged as shown above, and your Team’s Action is 12:

\begin{itemize}
    \item If the other team chooses 17, your team earns 12 points and the other team earns 17.
    \item If the other team chooses 15, your team earns 12 + 20 = 32 points and the other team earns 15 points.
    \item If the other team also chooses 12, both teams earn 12 points.
    \item If the other team chooses 19, your team earns 12 points and the other team earns 19 + 20 = 39 points.
\end{itemize}

\bigskip

\noindent \textbf{Important:} You will play three rounds, and the order of the boxes may differ from the example above and may change from round to round. At the beginning of each round, the screen will display the order of the boxes for that round. All participants will see the boxes in the same order as you do.

\bigskip

\noindent As described earlier, you will send your team partner a \emph{\textbf{Suggested Decision}} and a \emph{\textbf{Message}}. Remember to explain in the message the reasoning behind your suggested decision. After this information is exchanged, both of you enter your \emph{\textbf{Final Decision}}, from which the computer randomly chooses the \emph{\textbf{Team’s Action}}.

\bigskip

\noindent You will be paid the sum of all points you earn in the three rounds (plus the points you earn in the other part).

\bigskip

\noindent When you click the ``Ready'' button, you will start the first round of \textbf{Part II} of the experiment.

\newpage

\section{Classification Instructions}
\label{appendix:ra_instruction}

This appendix provides self-contained classification instructions for beauty contest games (\ref{appendix:ra_instruction_beauty}) and 11--20 games (\ref{appendix:ra_instruction_1120}). Note that Part 2 was distributed only after the classification of Part 1 had been completed.

\subsection{Beauty Contest Games}
\label{appendix:ra_instruction_beauty}

\noindent {\large\textbf{Part 1}}

\medskip

\noindent In the following we will describe the classification process for the analysis of our experiment. We, Wang, Ahmed, Unjitwattana, Xie, Fong, and Lin, assume that you are familiar with the level-$k$ model as it has been introduced by \cite{nagel1995unraveling} or represented by \cite{camerer2004cognitive}. However, in order to clarify potential questions of terminology, the appendix on page~4 reproduces the main features of the model in the terminology used in this document.

\bigskip

\noindent The classification proceeds in two steps, Part 1 and Part 2. You are now provided by us with the transcripts for Part 1. The transcripts differ in the amount of information about the decisions taken. Only in Part 2 will you see the suggested choices of the players that were made.

\bigskip

\noindent After your individual classification of Part 1, you will meet with your co-classifier to reconcile your classification. In this process, try to agree on common classifications if possible and note them in the third ``Reconciliation'' sheet for each game. If an agreement is not possible and you keep your initial individual classification, simply note nothing in the third sheet. After you finish this process for Part 1, you will hand in the individual classification sheets and the Reconciliation sheets, and we will provide you with the material for Part 2. If you have questions about the procedure at any point, simply write an email to us and we will clarify any point in an email to both of you.

\bigskip

\noindent For Part 1, follow the instructions of this booklet now. Read them entirely to get an overview and then start the classification. Please read the messages in each of three beauty contest games (coded as ``beauty 2/3,'' ``beauty 1/3'' and ``beauty 1/2'') and note for each player the minimum level of reasoning and the level-0 belief mean in the Part 1 excel sheet provided. Below you will find detailed instructions for classifying each player. Please limit yourself to making inferences only from what can clearly be derived from the message stated, i.e. do not try to think about what the player \emph{might have thought}.

\bigskip

\noindent\textbf{\emph{IMPORTANT: When you think that the information does not clearly lend itself to any inference, simply do not note any classification. Consequently, do not note anything if no statement has been made. Please note only those classifications for which you are certain. Make use of the comments space if you are not certain but still want to indicate a feature of the reasoning. Similarly, please comment if the statement exhibits some argument that does not fit the level-$k$ model as we present it here.}}

\bigskip
\bigskip

\noindent \textbf{Levels: Lower Bounds}

\medskip

\noindent For the lower bound on the level of reasoning, you should ask yourself: ``What is the minimum level of reasoning that this statement clearly exhibits?'' Once noted, you should be able to say to yourself: ``It seems impossible that the players' level of reasoning is below this number!''

\bigskip

\noindent Here we ask you to be very cautious with the classification, not giving away high levels easily. \textbf{Please only write down the highest lower-bound for which you are absolutely certain!} In Part 2 you will be asked to classify the upper bounds of the level of reasoning. That will be the time to be generous with the interpretation of the statements.

\bigskip

\noindent \textbf{Level 0}

\medskip

\noindent The player does not exhibit any strategic reasoning whatsoever. Different versions of this might be randomly chosen numbers, misunderstanding of the game structure or giving other non-strategic ``reasons'' for picking a number, e.g. taste. It is important that no best-responding to the others' play occurs. There could be considerations of what others might play, but without best responding to it. Examples: ``Let's use 50. This is the average between 0 and 100.'' ``It's random, so let's guess something.'' ``My favorite number is 74.''

\bigskip

\noindent \textbf{Level 1}

\medskip

\noindent This player best responds to something (e.g., by multiplying a number by $p$, the beauty contest game parameter). However, he does not realize that others will be strategic as well. Example (for $p = 2/3$): ``They will all go for a number of about 50–55. So, we should do something like 35.''

\bigskip

\noindent \textbf{Level 2}

\medskip

\noindent This player not only best responds (by calculating something times $p$), but also realizes that other players best respond as well. At level 2 the question about the extent of strategic reasoning of other players can come up. In the theory, this is reflected as a population belief on levels 0 and 1. Example: ``Thinking that others play 60, everybody will play 40. So, we should be more clever and play two thirds of 40.'' ``Some will just play 90, while others will think and play 60 in response. We should therefore play somewhere between 60 and 40.''

\bigskip

\noindent \textbf{Level 3}

\medskip

\noindent This player realizes that others could be level 2 and reacts by best responding to this as well. Put differently, he realizes that others realize that others best respond as well. As for all players above level 1, the extent of strategic reasoning by others is important for level 3 reasoners as well. In addition, they have to ask themselves how the level-2 players think about the distribution of level 1 and level 0 players.

\bigskip

\noindent \textbf{Level 4, 5, ...}

\medskip

\noindent The process continues to higher levels. More levels of best responses and higher orders of beliefs become relevant.

\bigskip

\noindent \textbf{Level 0 Belief Mean}
 
\medskip

\noindent If the message hints toward a value of the mean of the level-0 distribution, then indicate this value as level-0 belief mean. Remember, the level-0 belief mean is the starting point of the reasoning. Players of positive level start best responding on this number. Please note a number as level-0 belief mean only when this number is not logically derived through level reasoning or the like. If an interval is indicated, please note the average of the lower and upper bound. For example, ``I think the others play around 50–60.'' can be noted as a mean of 55. If only a qualitative statement is made about the level-0 belief mean, try to quantify it if possible. Otherwise, please write a short comment that indicates what is written down. Similarly, if a distribution is specified, please comment precisely on the relevant passage.

\bigskip

\noindent The literature usually assumes a mean of 50. Be reminded that this is only a common assumption which should not influence your considerations at this point.

\bigskip
\bigskip
\bigskip

\noindent {\large\textbf{Model and Terminology}}

\medskip

\noindent The level-$k$ model of bounded rationality assumes that players only think through a certain number ($k$) of best responses. The model has four main ingredients:

\bigskip

\noindent \textbf{Population Distribution}
 
\medskip

\noindent This distribution on $\mathbb{N}_0$ reflects the proportion of types with a certain level $k$.

\bigskip

\noindent \textbf{Level-0 Distribution}
 
\medskip

\noindent Also called level-0 action distribution. By definition, a level-0 player does not best respond. Hence, his actions are random to the game and distributed over the action space, which in our case is $\mathcal{A} = \{{0},{1},{2},...,{99},{100}\}$.

\bigskip

\noindent \textbf{Level-0 Belief}
 
\medskip

\noindent In the model, players with $k > 0$ best respond to what they believe the level-0 players play. Their level-0 belief might not be consistent with the level-0 distribution. For best responding, all that matters of the level-0 belief is the mean, which lies in $[0, 100]$. It is frequently assumed that the level-0 distribution and the level-0 belief are consistent, but for the classification this is irrelevant.

\bigskip

\noindent \textbf{Population Belief}
 
\medskip

\noindent Players do not expect other players to be of the same or a higher level of reasoning. For a level-$k$ player, the population belief is therefore defined on the set of levels strictly below~$k$. It follows that level-0 players have no defined belief, level-1 players have a trivial belief with full probability mass on $\{0\}$, level-2 players have a well defined belief on $\{{0}, {1}\}$. From level-3 higher order beliefs are relevant as level-3 players have to form a belief about level-2’s beliefs.

\bigskip
\bigskip
\bigskip

\noindent {\large\textbf{Part 2}}

\medskip

\noindent The classification proceeds in two steps, Part 1 and Part 2. You are now provided by us with the transcripts for Part 2. You can now see the suggested choices of the players that were made.

\bigskip

\noindent After your individual classification of Part 2, you will meet with your co-classifier to reconcile your classification. In this process, try to agree on common classifications if possible and note them in the third ``Reconciliation'' sheet. If an agreement is not possible and you keep your initial individual classification, simply note nothing in the third sheet. If you have questions about the procedure at any point, simply write an email to us and we will clarify any point in an email to both of you.

\bigskip

\noindent Please consider the information on each player in the three beauty contest games and note for each player the upper bound of level of reasoning, whether the equilibrium has been identified, and whether dominance reasoning has been applied in the Part 2 excel sheet provided. Below you will find detailed instructions for classifying each player. Please limit yourself to making inferences only from what can clearly be derived from the message and the action data, i.e. do not try to think about what the player \emph{might have thought}.

\bigskip

\noindent\textbf{\emph{Be reminded that when you think that the information does not clearly lend itself to any inference, simply do not note any classification. Consequently, do not note anything if no statement has been made! Please note only those classifications for which you are certain. Make use of the comments space if you are not certain but still want to indicate a feature of the reasoning. Similarly, please comment if the statement exhibits some argument that does not fit the level-$k$ model as we present it here.}}

\bigskip
\bigskip

\noindent \textbf{Levels: Upper Bounds}

\medskip

\noindent The upper bounds should give the maximum level of reasoning that could be interpreted into the statement. Therefore, you should ask yourself: ``What is the highest level of reasoning that can be underlying this statement?'' Once noted, you should be able to say: ``Although maybe not clearly communicated, this statement could be an expression of this level. If the player reasoned higher than this number, this was not expressed in the statement!''

\bigskip

\noindent Please refer to the level characterizations in Part 1 of the instructions.

\bigskip

\noindent \textbf{Type: Equilibrium Identification}

\medskip

\noindent With this dummy, you indicate whether the player realized that the unique equilibrium is 0. For this he has to mention the equilibrium action 0. It is not enough to describe a process of downward convergence. The equilibrium might be mentioned anywhere in the statement, so it is irrelevant whether he stops reasoning when he found the equilibrium strategies or whether he finds further arguments not to play 0. People will not necessarily use the word ``equilibrium,'' but they might describe that ``theoretically everybody should play 0'' or that ``the process will be going down to 0.''

\bigskip

\noindent Set the dummy to 0 if the equilibrium was not identified and to 1 if it was.

\bigskip

\noindent \textbf{Type: Dominance Reasoning}

\medskip

\noindent With this dummy, you indicate whether the reasoning applied the concept of dominance for the explanation. This is defined as involving iterative deletion of dominated strategies and randomly playing one of the remaining actions or best responding to a distribution over the partner's remaining action space. People will not necessarily use the word ``dominance,'' but they might describe that ``playing above 66 makes no sense.'' Note that ``everybody plays on average 50 so I should not play higher than 34'' is not a dominance reasoning, because it rules out strategies based on a distribution on the full action space. Dominance reasoning rules out first and plays a best response then.

\bigskip

\noindent Set the dummy to 1 if dominance was used in the argument and to 0 if it was not.

\subsection{11--20 Games}
\label{appendix:ra_instruction_1120}

\noindent {\large\textbf{Part 1}}

\medskip

\noindent In the following we will describe the classification process for the analysis of our experiment. We, Wang, Ahmed, Unjitwattana, Xie, Fong, and Lin, assume that you are familiar with the level-k model as it has been introduced by \cite{nagel1995unraveling} or represented by \cite{camerer2004cognitive}. However, in order to clarify potential questions of terminology, the appendix on page~4 reproduces the main features of the model in the terminology used in this document. To help you become familiar with the classification task, we also describe below the rules of the 11--20 games used in the experiment.

\bigskip
\bigskip

\noindent \textbf{11--20 Games}

\medskip

\noindent In an 11--20 game, two players simultaneously select one of 10 boxes arranged in a line, each containing an integer between 11 and 20. Players receive the number in the box they select, plus an additional 20 points if their selected box is exactly ``one to the left'' of the box chosen by the other player. Importantly, ``one to the left'' refers to the physical position in the line, not to the numerical value. 

\bigskip

\noindent The experiment consists of three different 11--20 games, which differ only in how the numbers are arranged in the line, in the following order:
\begin{itemize}
    \item 11--20 Extreme: \;\; 19 \;\; 18 \;\; 17 \;\; 16 \;\; 15 \;\; 14 \;\; 13 \;\; 12 \;\; 11 \;\; 20
    \item 11--20 Moderate: \;\; 14 \;\; 13 \;\; 12 \;\; 11 \;\; 19 \;\; 18 \;\; 17 \;\; 16 \;\; 15 \;\; 20
    \item 11--20 Basic: \;\; 11 \;\; 12 \;\; 13 \;\; 14 \;\; 15 \;\; 16 \;\; 17 \;\; 18 \;\; 19 \;\; 20
\end{itemize}

\bigskip
\bigskip

\noindent \textbf{Overview of the Classification Task}

\medskip

\noindent The classification proceeds in two steps, Part 1 and Part 2. You are now provided by us with the transcripts for Part 1. The transcripts differ in the amount of information about the decisions taken. Only in Part 2 will you see the suggested choices of the players that were made.

\bigskip

\noindent After your individual classification of Part 1, you will meet with your co-classifier to reconcile your classification. In this process, try to agree on common classifications if possible and note them in the third ``Reconciliation'' sheet for each game. If an agreement is not possible and you keep your initial individual classification, simply note nothing in the third sheet. After you finish this process for Part 1, you will hand in the individual classification sheets and the Reconciliation sheets, and we will provide you with the material for Part 2. If you have questions about the procedure at any point, simply write an email to us and we will clarify any point in an email to both of you.

\bigskip

\noindent For Part 1, follow the instructions of this booklet now. Read them entirely to get an overview and then start the classification. Please read the messages in each of the three 11--20 games (coded as ``11--20 Extreme,'' ``11--20 Moderate,'' and ``11--20 Basic'') and note for each player the minimum level of reasoning and the level-0 belief in the Part 1 excel sheet provided. Below you will find detailed instructions for classifying each player. Please limit yourself to making inferences only from what can clearly be derived from the message stated, i.e. do not try to think about what the player \emph{might have thought}.

\bigskip

\noindent\textbf{\emph{IMPORTANT: When you think that the information does not clearly lend itself to any inference, simply do not note any classification. Consequently, do not note anything if no statement has been made! Please note only those classifications for which you are certain. Make use of the comments space if you are not certain but still want to indicate a feature of the reasoning. Similarly, please comment if the statement exhibits some argument that does not fit the level-k model as we present it here.}}

\bigskip
\bigskip

\noindent \textbf{Levels: Lower Bounds}

\medskip

\noindent For the lower bound on the level of reasoning, you should ask yourself: ``What is the minimum level of reasoning that this statement clearly exhibits?'' Once noted, you should be able to say to yourself: ``It seems impossible that the players’ level of reasoning is below this number!''

\bigskip

\noindent Here we ask you to be very cautious with the classification, not giving away high levels easily. \textbf{Please only write down the highest lower-bound for which you are absolutely certain!} In Part 2 you will be asked to classify the upper bounds of the level of reasoning. That will be the time to be generous with the interpretation of the statements.

\newpage

\noindent \textbf{Level 0}

\medskip

\noindent The player does not exhibit any strategic reasoning whatsoever. Different versions of this might be randomly chosen numbers, misunderstanding of the game structure or giving other non-strategic ``reasons'' for picking a number, e.g. taste. It is important that no best-responding to the others' play occurs. There could be considerations of what others might play, but without best responding to it. Examples: ``Let’s use 15. It's in the middle.'' ``It's random, so let's guess something.'' ``My favorite number is 17.'' ``Let’s take 20 because it’s the biggest number.''

\bigskip

\noindent \textbf{Level 1}

\medskip

\noindent This player best responds to something (e.g. undercuts the opponent by choosing the box ``one to the left,'' the rule for getting the bonus in all three games). However, he does not realize that others will be strategic as well. Example (for Extreme): ``They will all go for 20. So, we should do 11.''

\bigskip

\noindent \textbf{Level 2}

\medskip

\noindent This player not only best responds (e.g. undercuts the opponent by choosing the box ``one to the left'') but also realizes that other players best respond as well. At level 2 the question about the extent of strategic reasoning of other players can come up. In the theory, this is reflected as a population belief on levels 0 and 1. Example (for Extreme): ``Thinking that others play 20, everybody will play 11. So, we should be cleverer and play one to the left of 11.'' ``Some will just play 20, while others will think and play 11 in response. We should therefore play either 11 or 12.''

\bigskip

\noindent \textbf{Level 3}

\medskip

\noindent This player realizes that others could be level 2 and reacts by best responding to this as well. Put differently, he realizes that others realize that others best respond as well. As for all players above level 1, the extent of strategic reasoning by others is important for level 3 reasoners as well. In addition, they have to ask themselves how the level-2 players think about the distribution of level 1 and level 0 players.

\bigskip

\noindent \textbf{Level 4, 5, ...}

\medskip

\noindent The process continues to higher levels. More levels of best responses and higher orders of beliefs become relevant.

\bigskip

\noindent \textbf{Level 0 Belief}
 
\medskip

\noindent If the message hints toward one or more specific integers in the support of the level-0 distribution, then indicate these integers as level-0 belief. Remember, the level-0 belief is the starting point of the reasoning. Players of positive level start best responding to this belief. Please note an integer or integers as level-0 belief only when these integers are not logically derived through level reasoning or the like. If several integers are indicated, please summarize the relevant level-0 belief as precisely as possible. For example, ``I think others play either 20 or 19'' can be noted as ``19, 20.'' If only a qualitative statement is made about the level-0 belief, try to quantify it if possible. Otherwise, please write a short comment that indicates what is written down. Similarly, if a distribution is specified, please comment precisely on the relevant passage.

\bigskip

\noindent The literature usually assumes a degenerate level-0 distribution at 20. Be reminded that this is only a common assumption which should not influence your considerations at this point.

\bigskip
\bigskip
\bigskip

\noindent {\large\textbf{Model and Terminology}}

\medskip

\noindent The level-$k$ model of bounded rationality assumes that players only think through a certain number ($k$) of best responses. The model has four main ingredients:

\bigskip

\noindent \textbf{Population Distribution}
 
\medskip

\noindent This distribution on $\mathbb{N}_0$ reflects the proportion of types with a certain level $k$.

\bigskip

\noindent \textbf{Level-0 Distribution}
 
\medskip

\noindent Also called level-0 action distribution. By definition, a level-0 player does not best respond. Hence, his actions are random to the game and distributed over the action space $\mathcal{A} = \{11,12,...,19,20\}$.

\bigskip

\noindent \textbf{Level-0 Belief}
 
\medskip

\noindent In the model, players with $k > 0$ best respond to what they believe the level-0 players play. Their level-0 belief might not be consistent with the level-0 distribution. It is frequently assumed that the level-0 distribution and the level-0 belief are consistent, but for the classification this is irrelevant.

\bigskip

\noindent \textbf{Population Belief}
 
\medskip

\noindent Players do not expect other players to be of the same or a higher level of reasoning. For a level-$k$ player, the population belief is therefore defined on the set of levels strictly below~$k$. It follows that level-0 players have no defined belief, level-1 players have a trivial belief with full probability mass on $\{0\}$, level-2 players have a well defined belief on $\{{0}, {1}\}$. From level-3 higher order beliefs are relevant as level-3 players have to form a belief about level-2’s beliefs. 

\bigskip
\bigskip
\bigskip

\noindent {\large\textbf{Part 2}}

\medskip

\noindent The classification proceeds in two steps, Part 1 and Part 2. You are now provided by us with the transcripts for Part 2. You can now see the suggested choices of the players that were made. 

\bigskip

\noindent After your individual classification of Part 2, you will meet with your co-classifier to reconcile your classification. In this process, try to agree on common classifications if possible and note them in the third ``Reconciliation'' sheet for each game. If an agreement is not possible and you keep your initial individual classification, simply note nothing in the third sheet. If you have questions about the procedure at any point, simply write an email to us and we will clarify any point in an email to both of you.

\bigskip

\noindent Please consider the information on each player in the three 11--20 games and note for each player the upper bound of level of reasoning, whether security has been applied, and whether undercut aversion has been applied in the Part 2 excel sheet provided. Below you will find detailed instructions for classifying each player. Please limit yourself to making inferences only from what can clearly be derived from the message and the action data, i.e. do not try to think about what the player \emph{might have thought}.

\bigskip

\noindent\textbf{\emph{Be reminded that when you think that the information does not clearly lend itself to any inference, simply do not note any classification. Consequently, do not note anything if no statement has been made! Please note only those classifications for which you are certain. Make use of the comments space if you are not certain but still want to indicate a feature of the reasoning. Similarly, please comment if the statement exhibits some argument that does not fit the level-$k$ model as we present it here.}}

\bigskip
\bigskip

\noindent \textbf{Levels: Upper Bounds}

\medskip

\noindent The upper bounds should give the maximum level of reasoning that could be interpreted into the statement. Therefore, you should ask yourself: ``What is the highest level of reasoning that can be underlying this statement?'' Once noted, you should be able to say: ``Although maybe not clearly communicated, this statement could be an expression of this level. If the player reasoned higher than this number, this was not expressed in the statement!''

\bigskip

\noindent Please refer to the level characterizations in Part 1 of the instructions.

\bigskip

\noindent \textbf{Type: Security}

\medskip

\noindent With this dummy, you indicate whether the player takes securing a safe payoff into account when reasoning about which numbers to choose. For this, the player's statement has to reflect a concern about securing a safe payoff (e.g. by considering relatively high numbers such as 20), regardless of whether undercutting logic is also present. Players will not necessarily use the word ``safe'' or ``risky,'' but they might describe that ``let’s just take 20 and lock in the points,'' or that ``if we try to undercut them by playing a small number but guess wrong, we might lose a lot.''

\bigskip

\noindent Set the dummy to 1 if security was used in the argument and to 0 if it was not.

\newpage

\noindent \textbf{Type: Undercut Aversion}

\medskip

\noindent With this dummy, you indicate whether the player takes preventing the other team from picking ``one to the left'' and claiming the 20 bonus into account when reasoning about which numbers to choose. For this, the player's statement has to express the motive of not wanting the other team to get the bonus. A bare choice of the leftmost box is not enough. A qualifying example is: ``Let’s take the leftmost box so the other team can't sit to our left and grab the bonus off us.'' People will not necessarily use the words ``undercut'' or ``bonus,'' but they might describe that ``this choice is unbeatable.''

\bigskip

\noindent Set the dummy to 1 if undercut aversion was used in the argument and to 0 if it was not.

\bigskip

\noindent\textbf{\emph{Important: The rationales of Security and Undercut Aversion are separate from, and not mutually exclusive with, the logic of undercutting. Players may rely on multiple rationales simultaneously to justify their suggested choices. Code each rationale independently whenever it is explicitly expressed.}}

\newpage

\section{Screenshots}
\label{appendix:screenshots}

\begin{figure}[htbp!]
    \centering
    \includegraphics[width=\linewidth]{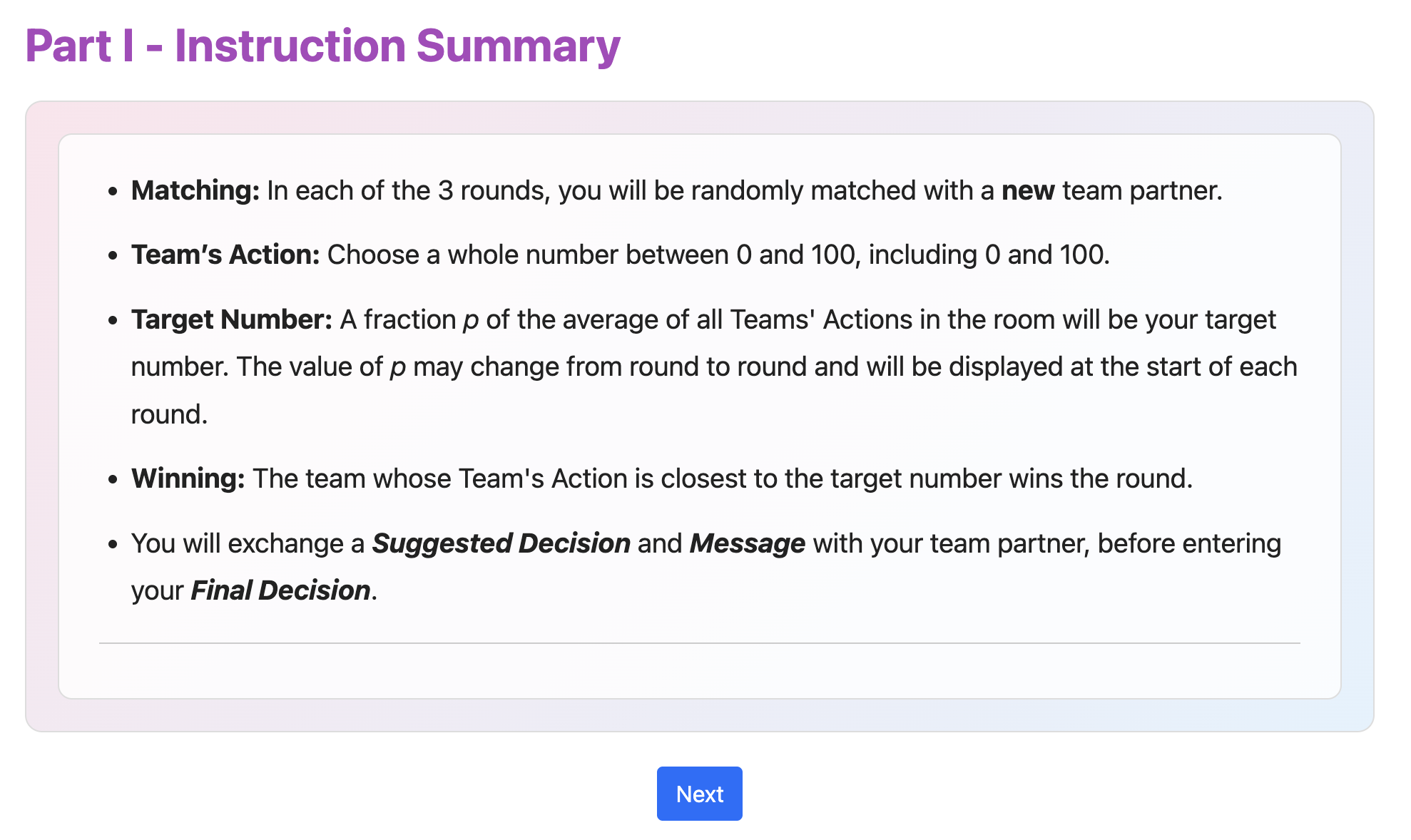}
    \caption{Instruction Summary of the Beauty Contest Game}
    \label{fig:beauty_screen_1}
\end{figure}

\begin{figure}[htbp!]
    \centering
    \includegraphics[width=\linewidth]{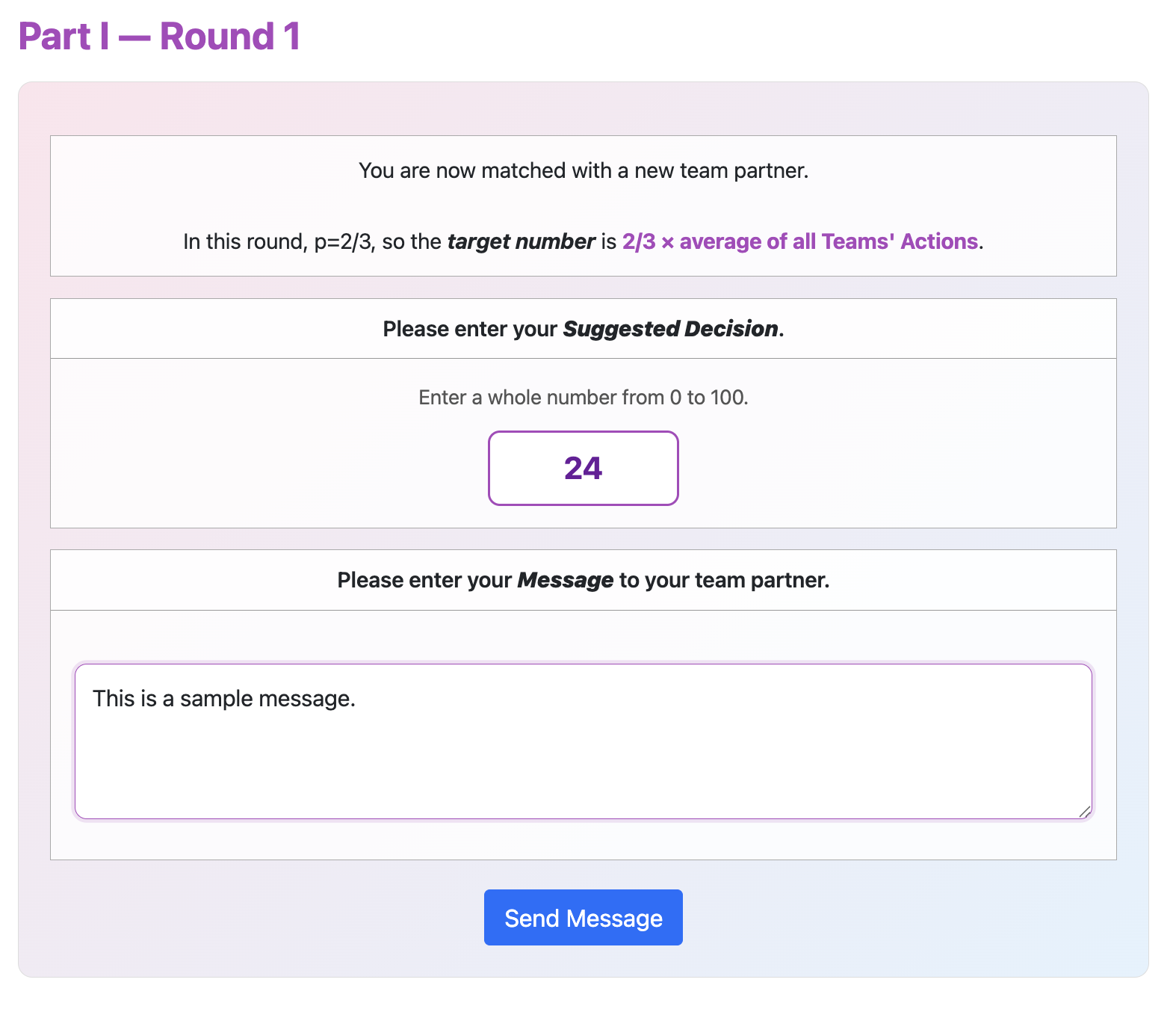}
    \caption{Messaging Stage of the Beauty Contest Game}
    \label{fig:beauty_screen_2}
\end{figure}

\begin{figure}[htbp!]
    \centering
    \includegraphics[width=\linewidth]{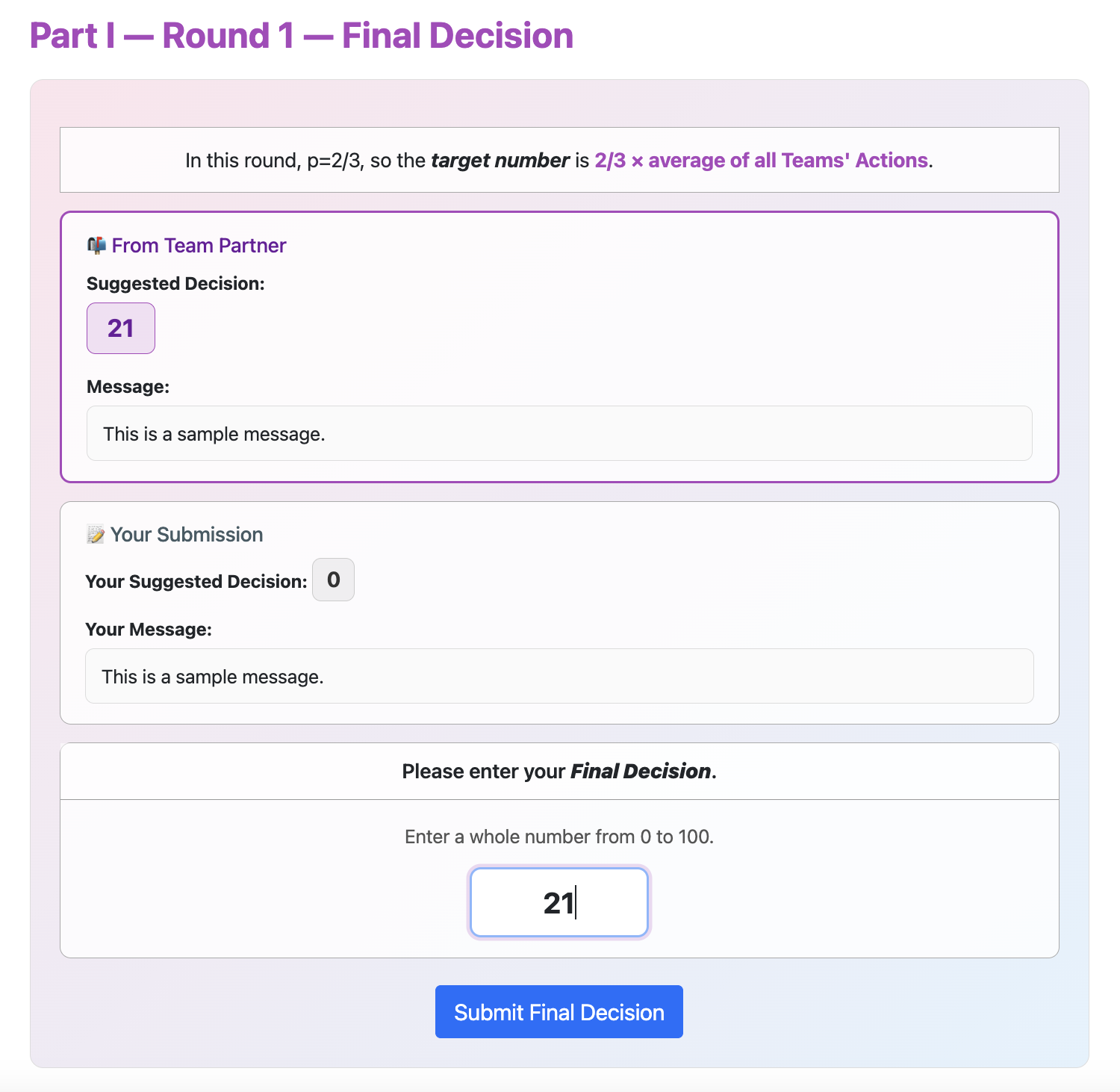}
    \caption{Decision Stage of the Beauty Contest Game}
    \label{fig:beauty_screen_3}
\end{figure}

\begin{figure}[htbp!]
    \centering
    \includegraphics[width=\linewidth]{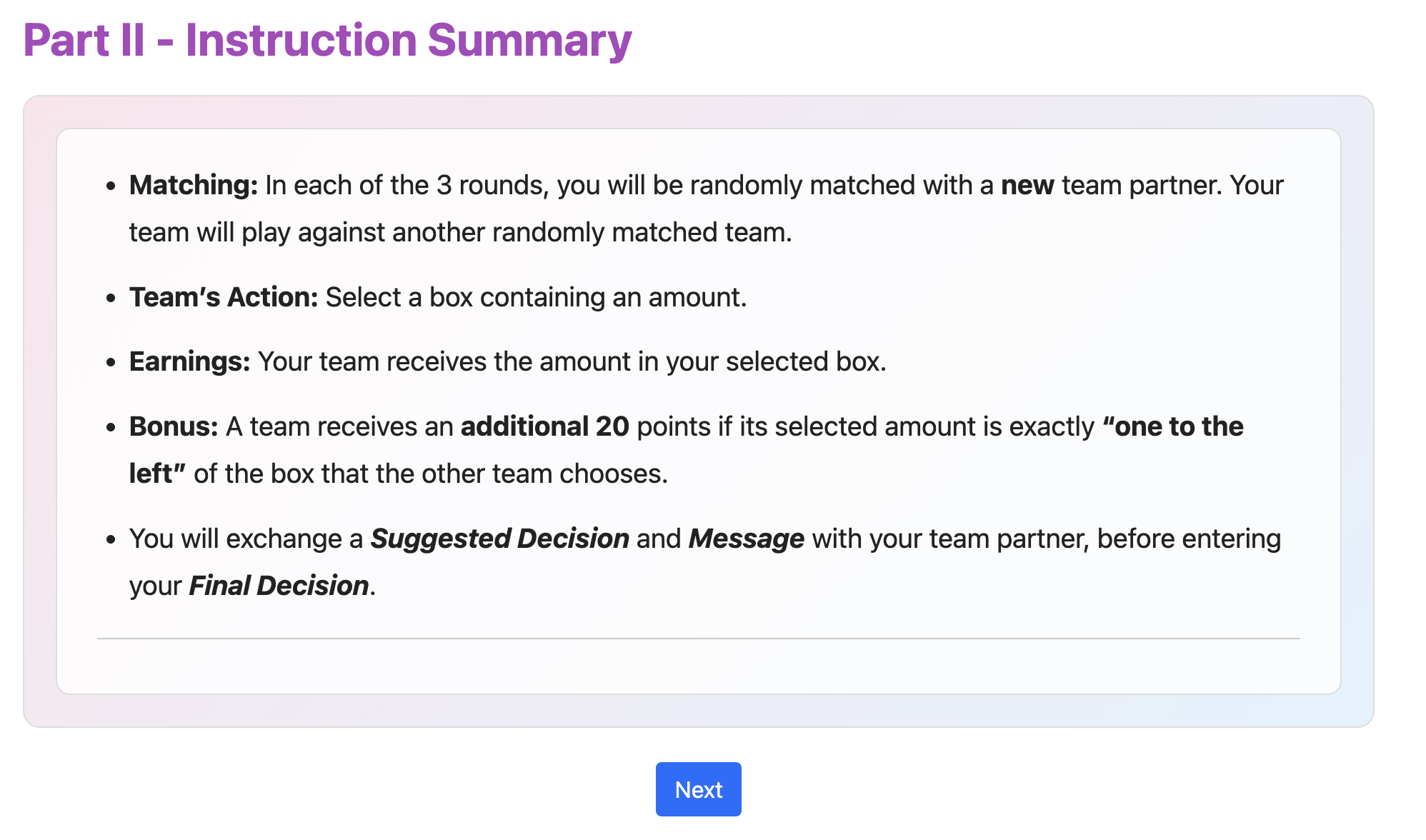}
    \caption{Instruction Summary of the 11--20 Game}
    \label{fig:11_20_1}
\end{figure}

\begin{figure}[htbp!]
    \centering
    \includegraphics[width=\linewidth]{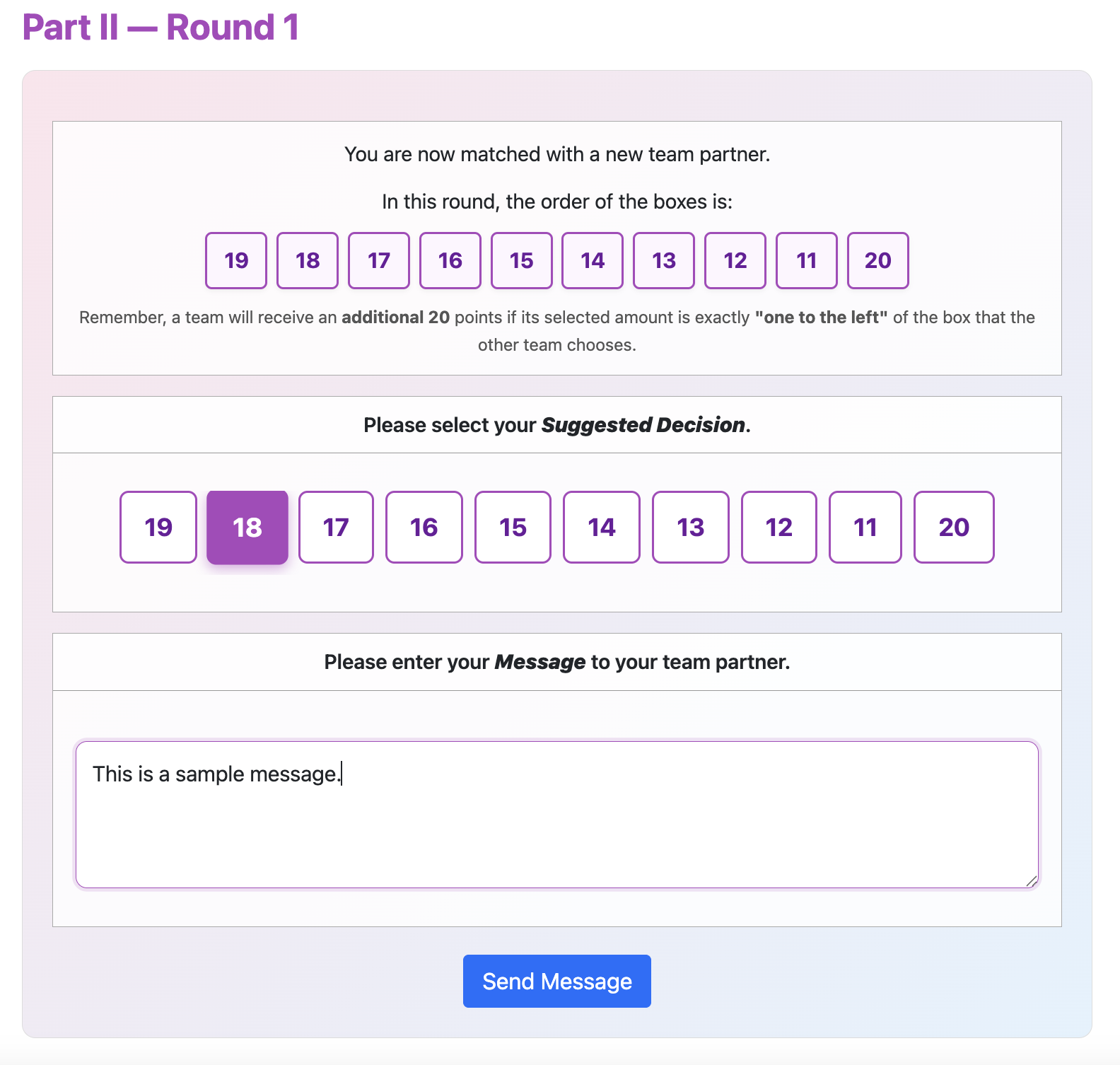}
    \caption{Messaging Stage of the 11--20 Game}
    \label{fig:11_20_2}
\end{figure}

\begin{figure}[htbp!]
    \centering
    \includegraphics[width=\linewidth]{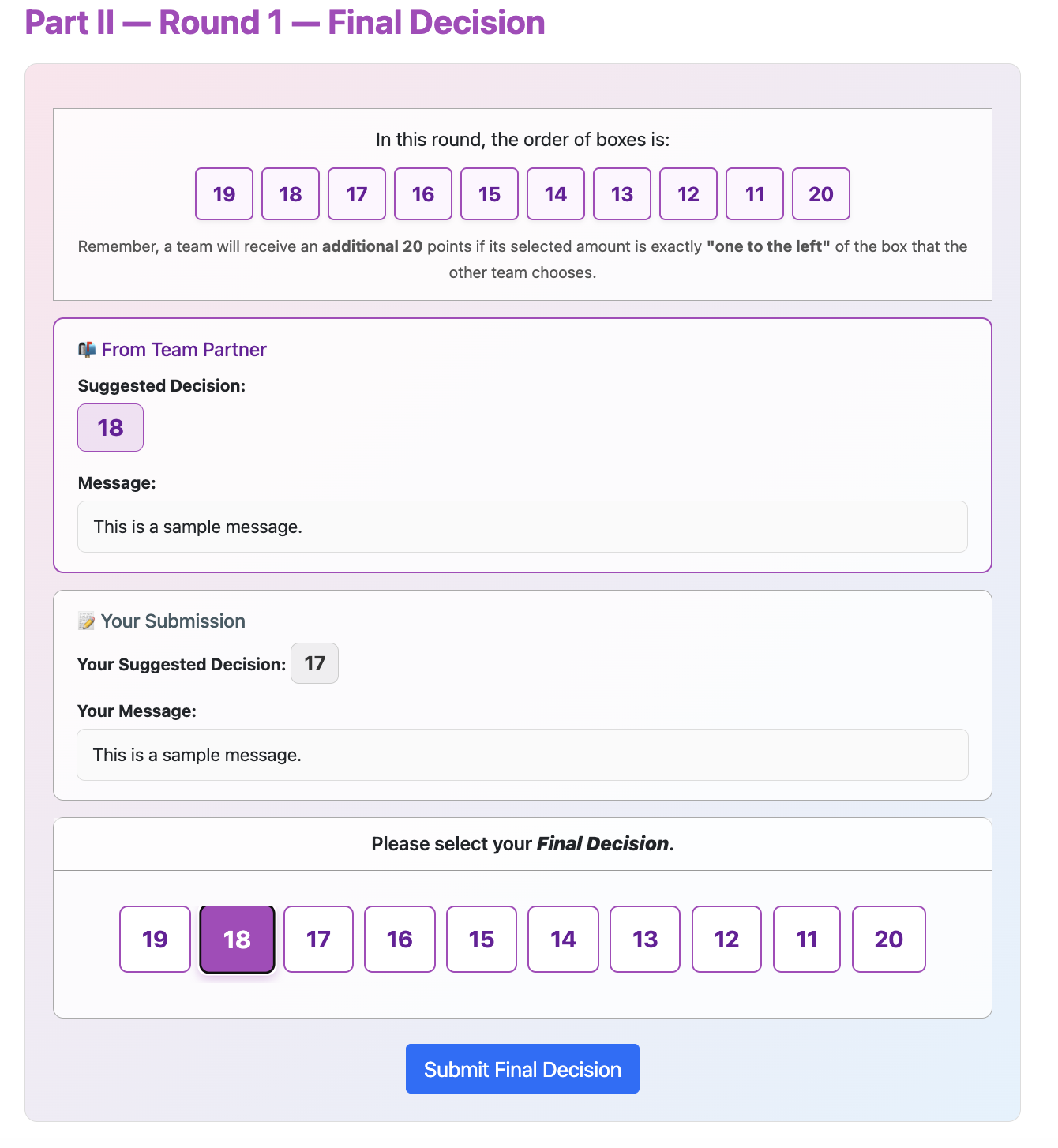}
    \caption{Decision Stage of the 11--20 Game}
    \label{fig:11_20_3}
\end{figure}

\begin{figure}[htbp!]
    \centering
    \includegraphics[width=\linewidth]{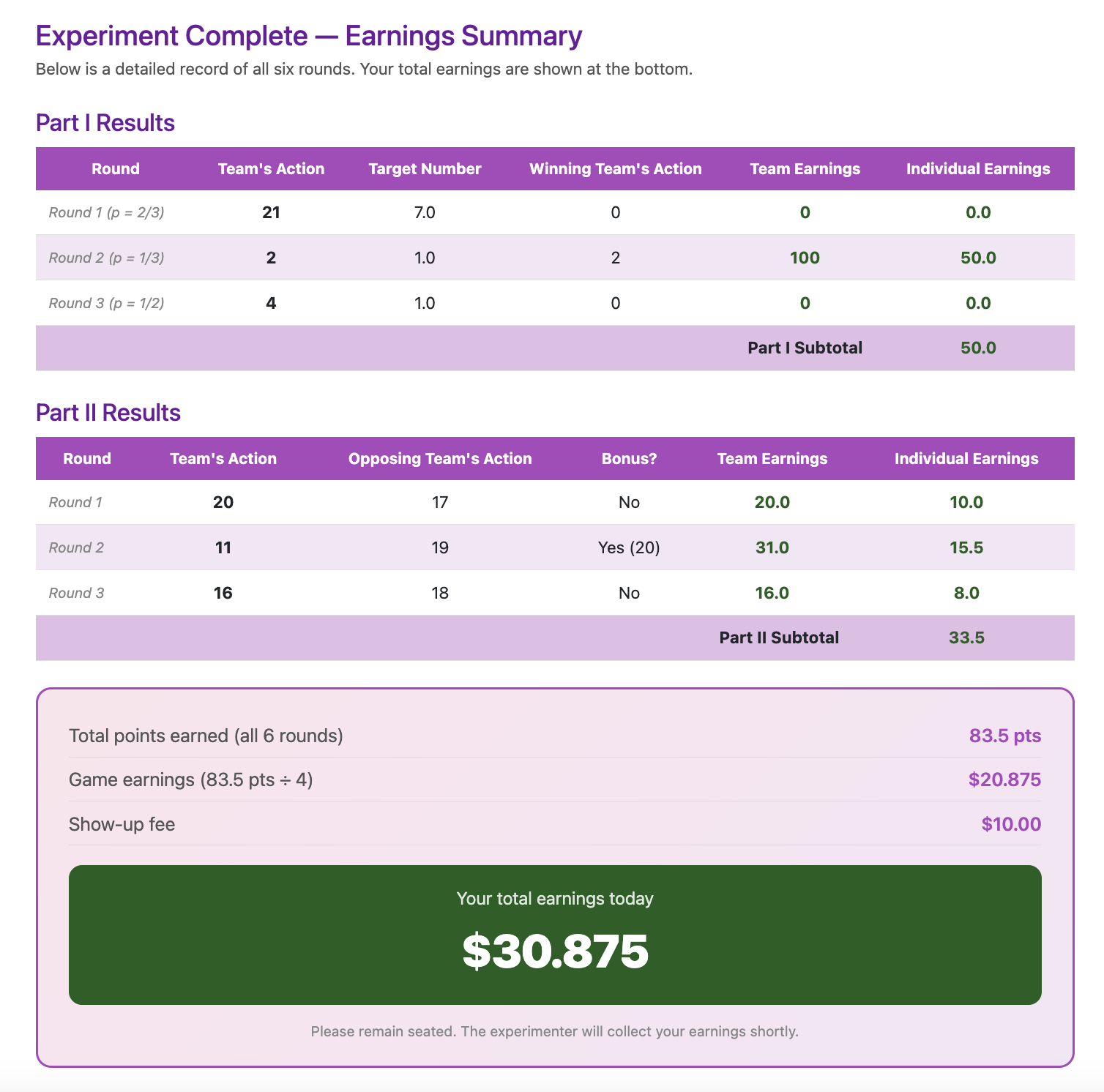}
    \caption{Feedback and Earnings Summary}
    \label{fig:feedback_screen}
\end{figure}

In this appendix, we provide screenshots of the experimental program 
for the treatment in which subjects play the beauty contest games first, 
followed by the 11--20 games.\footnote{The screenshots for the other treatment order are identical, except that Parts One and Two are swapped.} Figures \ref{fig:beauty_screen_1} to 
\ref{fig:beauty_screen_3} present sample screenshots from the first round 
of the beauty contest game, where $p = 2/3$. In rounds 2 and 3, the multipliers are $1/3$ 
and $1/2$, respectively. Figures \ref{fig:11_20_1} to \ref{fig:11_20_3} present sample screenshots from the first round of 
the re-ordered 11--20 game. The configuration of the first round corresponds to the extreme 
version in GLZ, while rounds 2 and 3 correspond to the moderate and 
baseline versions, respectively. Finally, Figure \ref{fig:feedback_screen} presents a sample screenshot of the feedback screen shown at the end of the experiment.

\end{document}